\documentclass[]{aastex631}
\usepackage{natbib}
\usepackage{graphicx}
\usepackage{multirow}
\usepackage{subfigure} 
\usepackage{amsmath}	
\usepackage{amssymb}	
\usepackage{bm}		
\usepackage{color}
\usepackage{float}
\usepackage{chngpage}
\usepackage{longtable}

\DeclareMathSizes{10}{10}{7}{3}

\received{-}
\revised{-}
\accepted{-}
\submitjournal{ApJS}

\shorttitle{The Statistic Analysis of GeV Spectral Breaks in Bright Gamma-Ray FSRQs}
\shortauthors{Zhu, Chen, Zhang}

\begin{document}

\title{The Statistic Analysis of GeV Spectral Breaks in Bright Gamma-Ray Flat Spectrum Radio Quasars}

\correspondingauthor{L. Zhang}\email{lizhang@ynu.edu.cn}

\author[0000-0002-3132-1507]{K.R.Zhu}
\affiliation{Department of Astronomy, Key Laboratory of Astroparticle Physics of Yunnan Province, Yunnan University, Kunming, 650091, People's Republic of China}

\author[0000-0001-5681-6939]{J.M. Chen}
\affiliation{Department of Astronomy, Key Laboratory of Astroparticle Physics of Yunnan Province, Yunnan University, Kunming, 650091, People's Republic of China}

\author{L. Zhang}
\affiliation{Department of Astronomy, Key Laboratory of Astroparticle Physics of Yunnan Province, Yunnan University, Kunming, 650091, People's Republic of China}

\begin{abstract}
We present the statistic results of GeV spectral breaks of bright gamma-ray flat-spectrum radio quasars (FSRQs) in the energy range of 0.1-10 GeV based on New Pass 8 date of the Large Area Telescope abroad on Fermi Gamma-ray Space Telescope. We have fitted the 15-year average gamma-ray spectra of 755 FSRQs by using both a broken power-law (BPL) and the Logarithmic Parabolas (LP) models, and obtained 87 bright gamma-ray FSRQs with their integrated photon fluxes greater than $2.16\times 10^{-8}$ cm$^{-2}$ s$^{-1}$. From our results, the FSRQ population shows similar preferences for both the BPL and LP models in gamma-ray spectral fitting and clustering analysis suggests that BPL-preferred and LP-preferred FSRQs belong to the same category. Our results indicate that GeV spectral breaks in bright gamma-ray FSRQs are located at $\rm 2.90\pm1.92$ GeV in the rest frame, and the observed change in photon index is $\rm \Delta \gamma =0.45 \pm 0.19$, which is consistent with the expected value for a cooling break of electrons scattering seed photons.

\end{abstract}
\keywords{Active galactic nuclei; Gamma-ray sources; Blazars; Flat-spectrum radio quasars; data analysis}

\section{Introduction}\label{sec:intro}

Flat-spectrum radio quasars (FSRQs) are a subclass of blazars, characterized by strong spectral emission lines (equivalent width $\rm  >5\AA$, \citealt{1991ApJ...374..431S}). Observed multi-wavelength spectral energy distributions (SEDs) of FSRQs exhibits a double-peaked structure. Usually, the observed SEDs can be explained in the leptonic model framework. In this case, the low-energy peak is generally attributed to originate from synchrotron emission of relativistic electrons in the jet, and high-energy peak is attributed to the inverse Compton (IC) scattering of seed photons by jet electrons, where the seed photons may originate from the jet's synchrotron emission (synchrotron self-Compton, or SSC; e.g. \citealt{1981ApJ...243..700K,1985ApJ...298..114M,1989ApJ...340..181G,1992ApJ...397L...5M,2016MNRAS.457.3535Z}) and external photon fields (external Compton, or EC; e.g. \citealt{1993ApJ...416..458D,1994ApJ...421..153S,1998MNRAS.299..433F,2009MNRAS.397..985G,2021MNRAS.500.5297A}), including the accretion disk \citep{1993ApJ...416..458D,2013MNRAS.429.1189P}, broad-line region (BLR, e.g. \citealt{1994ApJ...421..153S,1996MNRAS.280...67G}), and dust torus (DT, e.g. \citealt{2000ApJ...545..107B,2002A&A...386..415A}). 

The GeV spectral breaks of Flat-spectrum radio quasars (FSRQs) have been identified shortly after the Large Area Telescope (LAT) of Fermi Gamma-ray Space Telescope (Fermi-LAT) \citep{2009ApJ...697.1071A} was launched. Early observations by Fermi-LAT revealed that the GeV spectra of some FSRQs cannot be described by a simple Power Law (PL) or smooth curves, but rather exhibit a broken power-law (BPL) distribution \citep{2010ApJ...710.1271A}. For example, the Fermi-LAT team found a GeV break at $\sim$ 2 GeV in the bright FSRQ 3C 454.3, with the break energy remaining relatively stable across different activity phases \citep{2009ApJ...699..817A, 2010ApJ...721.1383A, 2011ApJ...733...19T}. Additionally, significant GeV spectral breaks within the 1-10 GeV energy range have also been detected in a dozen of bright gamma-ray blazars, primarily FSRQs, with a few being low-synchrotron-peaked BL Lacs \citep{2010ApJ...710.1271A, 2010ApJ...717L.118P, 2012ApJ...761....2H}. However, previous studies mainly focused on identifying the GeV spectral break and investigating its origin \citep[e.g.][]{2010ApJ...710.1271A,2010ApJ...717L.118P, 2011ApJ...733...19T, 2012ApJ...761....2H, 2014ApJ...794....8S, 2015MNRAS.449.2901K, 2015MmSAI..86...23B}. Due to limited exposure time, it was restricted to a very small number of bright gamma-ray sources. Moreover, because of the limitations in spectral resolution, there is significant uncertainty in fitting the GeV spectra. Extending the energy range to tens or even hundreds of GeV introduces complex processes such as extragalactic background light absorption, which poses challenges in investigating the origin of the GeV spectral break.

At present, however, the origin of the GeV spectral breaks in FSRQs remains an open issue. There are different physics explanations for the GeV spectral break of a given FSRQ. For example, it would come from the absorption of gamma-rays via photon-photon pair production by BLR photons \citep{2010ApJ...717L.118P, 2012ApJ...761....2H} or intrinsic electron spectrum breaks \citep{2009ApJ...699..817A, 2010ApJ...721.1383A,2010ApJ...714L.303F, 2021MNRAS.502.5875K}; it can also be produced by the Klein Nishina (KN) effect when external photons are scattered by jet electrons \citep{2013ApJ...771L...4C, 2021MNRAS.502.5875K} and/or scattering of mixed external photon fields by jet electrons (e.g., \citealt{2010ApJ...714L.303F,2013ApJ...771L...4C,2016A&A...589A..96H, 2021MNRAS.502.5875K}). 
Therefore, to better understand the origin of GeV spectral breaks of FSRQs, a comprehensive study with a larger sample is required.

Since its launch in 2008, Fermi-LAT has been conducting continuous surveys for over 15 years. The long-period survey by Fermi-LAT, along with optimizations in background models and data processing procedures, has facilitated a systematic investigation of GeV spectral breaks. In this work, we present a systematic investigation of GeV spectral breaks in bright gamma-ray FSRQs. Using the new Pass 8 data (P8R3) from Fermi-LAT, we fit the 15-year average gamma-ray spectra of all FSRQs using both a broken power-law (BPL) and the LogParabola (LP) models. From these, we select bright samples, perform statistical analysis on the fitting results, and infer the origins of the GeV spectral breaks.

The structure of this paper is as follows: Section 2 provides a brief overview of the sample selection and the Fermi analysis process. It presents the fitting results for both the LP and BPL models and compares the goodness-of-fit between the two models. Section 3 discusses the GeV behavior of the FSRQ population, compares these results with previous works, and explores the origin of GeV spectral breaks. Finally, Section 4 offers conclusions and discussions.

In this paper, following cosmological parameters are used: the Hubble constant $\rm H_{0}$ is set to $\rm 75\, km\, s^{-1}\, Mpc^{-1}$, the matter energy density $\rm \Omega_M$ is 0.27, the radiation energy density $\rm \Omega_rz$ is 0, and the dimensionless cosmological constant $\Omega_{\Lambda}$ is 0.73. In algorithms involving random processes, the random seed is initialized as ``123".

\section{ Fermi Spectrum Analysis of FSRQs} \label{sec:spectrum analysis}
The newest AGN catalog of Fermi-LAT, 4LAC-DR3, reported 3816 AGNs detected within 2008-2020 observations \citep{2022ApJS..263...24A}. From the 3407 high Galactic latitude samples\footnote{\url{https://fermi.gsfc.nasa.gov/ssc/data/access/lat/4LACDR3/table-4LAC-DR3-h.fits}}, 755 FSRQs with a known redshift are selected. 

\subsection{Default model fitting}

\begin{table*}[ht]
\begin{center}
\caption{Comparison of fitting parameters between bright and dark FSRQs}
\label{tab:1}
\setlength{\LTleft}{0 cm} \setlength{\LTright}{0 cm} 
\begin{adjustwidth}{-1.2cm}{-5cm}
\resizebox{19cm}{!}{
\begin{tabular}{cccc}
\hline
Parameter & Description&  Bright FSRQs & Dark FSRQs \\
\hline
Counts & Source count & 87 & 409 \\
$F_\gamma$& The 15-year integrated photon flux over 0.1-10 GeV  & 0.22 - 12.26 & 0.001- 0.76 \\
TS & TS value over 0.1-10 GeV &  5177 - 629618 & 4 - 4956 \\
Spectral type & Spectral models used for fitting gamma-ray spectra & LP (87) & LP (250) / PL (159) \\
Relative error & Average relative error of fitting parameters &  7.5\% & 33.8\% \\
$\rm R(5\sigma)$ & Average proportion of energy bins reaching 5$\sigma$ significance over 0.1-10 GeV &  99.9\% & 53.9\% \\
\hline
\end{tabular}}
\end{adjustwidth}
\end{center}
{\footnotesize{{Note. 
 \\
$F_\gamma$ is in unit of $\rm 10^{-7}\ cm^{-2}s^{-1}$ \\
}}}
\end{table*}

The Fermi-LAT collaboration has analyzed the 14-year average gamma-ray spectrum of 755 FSRQs from 2008 to 2022 in the energy range from 50 MeV to 1 TeV \citep{2023arXiv230712546B}. In their work, the GeV gamma-ray spectra of bright FSRQs are characterized using the Logarithmic Parabolas model (LP)\footnote{See \url{https://fermi.gsfc.nasa.gov/ssc/data/analysis/scitools/source_models.html2} for more details.}
\begin{equation}\label{eq1}
\frac{d N}{d E}=N_0\left(\frac{E}{E_0}\right)^{-\left(\alpha+\beta \ln \left(E / E_0\right)\right)} \;,
\end{equation}
or the Super Exponential Cutoff Power Law model (PLEC4)
\begin{equation}\label{eq2}
\frac{d N}{d E}= \begin{cases}N_0\left(\frac{E}{E_0}\right)^{-\gamma_0-\frac{d}{2} \ln \frac{E}{E_0}-\frac{d b}{6} \ln ^2 \frac{E}{E_0}-\frac{d b^2}{24} \ln ^3 \frac{E}{E_0}}, & \text { if }\left|b \ln \frac{E}{E_0}\right|<1 e^{-2} \\ N_0\left(\frac{E}{E_0}\right)^{-\gamma_0+d / b} \exp \left(\frac{d}{b^2}\left(1-\left(\frac{E}{E_0}\right)^b\right)\right) & \text { otherwise }\end{cases}.
\end{equation}
 The GeV gamma-ray spectra of dark FSRQs are characterized by  the simple Power Law model (PL):
\begin{equation}\label{eq3}
\frac{d N}{d E}=N_0\left(\frac{E}{E_0}\right)^{-\gamma} \text {.}
\end{equation}

Using these spectral models, we first perform a standard binned likelihood analysis of the gamma-ray spectra of 755 FSRQs over the 0.1-10 GeV range using 15 years of Fermi-LAT data. Since the PLEC4 model does not exhibit a cutoff below 10 GeV, making it similar to the LP model, we have also used the LP model to fit four PLEC4 samples (3C 454.3, CTA 102, 3C 279, and PKS 1424-41) to maintain consistency in the work. In summary, out of 755 samples, 437 sources were fitted using the LP model, while the remaining 318 samples were fitted using the PL model to extract their spectra.

An analysis pipeline was built with the \emph{Fermitools} (version v11r5p3) and \emph{Fermipy} \citep{2017ICRC...35..824W}. For each FSRQ, 15 years of Fermi-LAT P8R3 data were gathered from the Fermi Science Support Center (FSSC), covering the period from October 1st, 2008, to October 1st, 2023, in Greenwich Mean Time (MET from 244512001 to 718711205). The analysis models were constructed within a $15^{\circ}$ Region of Interest (ROI). Photon selection followed default criteria `(DATA$\_$ QUAL$>$0)$\&\&$(LAT$\_ $CONFIG==1)', with a zenith angle of  $90^{\circ}$. The corresponding instrument response function is \emph{P8R3$\_$SOURCE$\_$V3}, and the Galactic interstellar emission model is \emph{gll$\_$iem$\_$v07}.

The significance of a source was evaluated using the Test Statistics (TS) test, $\rm TS=2(ln{\mathcal L}_0-ln{\mathcal L}_1)$, where $\rm {\mathcal L}_0$ and $\rm {\mathcal L}_1$ are the maximum values of the likelihood function of the model including the target source and without the target source, respectively (e.g., \citealt{1996ApJ...461..396M}).  The square root of the TS value is approximately equal to the significance of the target source in units of $\sigma$ over the analysis energy band.

In the 755 FSRQs, due to the longer exposure times we employed, 259 sources were not successfully fitted, including:
\begin{itemize}
\item [i.] Cases where the Fermi fitting failed to converge (17 sources).
\item [ii.] Cases where, although the Fermi fitting converged, key parameters such as spectral index $\gamma$, and spectral curvature $\beta$ did not find optimal solutions within the predetermined reasonable range (242 sources).
\end{itemize}

After filtering the sample set, we found that 87 of the 755 FSRQs have an integrated photon flux greater than $2.16 \times 10^{-8}$ ph cm$^{-2}$ s$^{-1}$ in the 0.1-10 GeV range and a TS value greater than 5000 (gamma-ray significance higher than $70\sigma$) and their gamma-ray spectra are well fitted (we call them as bright gamma-ray FSRQs). We compared the fitting results of these 87 bright gamma-ray FSRQs with those of the remaining dark gamma-ray FSRQs, and the main fitting parameters are shown in Table \ref{tab:1}.

The average integrated gamma-ray photon flux of the dark FSRQs is about an order of magnitude lower than that of the bright sources. As a result, the gamma-ray spectral fits for the dark sources are poor. Specifically, the relative error of the fitting parameters for the dark sources averages 33.8\%, whereas it is only 7.5\% for the bright sources. Additionally, only about half of the energy bins in the 0.1 to 10 GeV range for the dark sources reach 5$\sigma$ significance, while nearly all energy bins for the bright sources have high significance. For the accuracy of our study, we will focus on these 87 bright sources for subsequent analysis \footnote{The fitting results of the dark sources are available online as FITS tables}.

In the 87 bright gamma-ray FSRQs, none of the sources is fitted with a PL spectrum; all are fitted with the LP model. During the fitting process, the Pivot Energy ($\rm E_0$ of Equation \ref{eq1}) was fixed at the values given in \cite{2023arXiv230712546B}; the optimal spectral index ($\alpha$ of Equation \ref{eq1}), spectral curvature ($ \beta$ of Equation \ref{eq1}), and differential flux at $ E_0$ ($N_0$ of Equation \ref{eq1}) were obtained from spectral fitting, with 1-$\sigma$ errors provided. Additionally, the integrated photon flux $ F_\gamma$ of the target source over 15 years from 0.1 to 10 GeV was fitted. The gamma-ray luminosity was calculated as $L_\gamma=4 \pi d_{\mathrm{L}}^2 S_\gamma$, where $d_{\mathrm{L}}$ is the luminosity distance, and $S_\gamma$ denotes the integrated energy flux obtained from the fitting in the 0.1-10 GeV range. All LP fitted results are tabulated in Table \ref{tab:2}.

The spectral index ranges from 1.86 to 2.81, with an average value of $\langle \alpha \rangle =2.24\pm 0.16$, and the spectral curvature ranges from -0.02 to 0.22, with an average value of $ \langle \beta \rangle =0.08\pm 0.04$. The fitting results are similar to those in 4FGL-DR4. The LP best-fitting lines in the $ E- E^2 dN/dE$ frame are shown as red lines in Figure \ref{fig:fig1}, and the corresponding 1-$\sigma$ confidence regions are marked in red.

\begin{table}
\setlength{\abovecaptionskip}{0 cm}
\setlength{\belowcaptionskip}{0.1cm}
\begin{center}
\caption{LP fitting results of 87 bright FSRQs}
\label{tab:2}
\centering
\small
\setlength{\LTleft}{0 cm} \setlength{\LTright}{0 cm} 
\begin{adjustwidth}{-2.2cm}{-5cm}
\resizebox{20cm}{!}{
\begin{tabular}{cccccccccc} \hline
\renewcommand\arraystretch{2}
\centering
{Source Name}&FGL Name & TS& $N_0$ & $E_0$ & $\alpha$ & $\beta$  &$\rm F_\gamma$&$\rm log_{10}(L_\gamma)$\\
\normalsize(1) & \normalsize(2) & \normalsize(3) &\normalsize(4) & \normalsize(5) &  \normalsize(6) &  \normalsize(7)    &  \normalsize(8)   &  \normalsize(9)   \\

\hline 
3C 454.3                    	&	4FGL J2253.9+1609	&	629618 	&	11.77 	$\pm$	0.04 	&	0.91 	&	2.39 	$\pm$	0.00 	&	0.11 	$\pm$	0.00 	&	12.26 	$\pm$	0.06 	&	48.17 	\\
PKS 1510-089                	&	4FGL J1512.8-0906	&	192398 	&	5.88 	$\pm$	0.03 	&	0.86 	&	2.37 	$\pm$	0.00 	&	0.03 	$\pm$	0.00 	&	6.54 	$\pm$	0.05 	&	47.06 	\\
3C 279                      	&	4FGL J1256.1-0547	&	259149 	&	5.35 	$\pm$	0.03 	&	0.98 	&	2.27 	$\pm$	0.00 	&	0.07 	$\pm$	0.00 	&	6.25 	$\pm$	0.03 	&	47.47 	\\
CTA 102                     	&	4FGL J2232.6+1143	&	264103 	&	4.00 	$\pm$	0.02 	&	1.10 	&	2.30 	$\pm$	0.01 	&	0.07 	$\pm$	0.00 	&	6.09 	$\pm$	0.04 	&	48.08 	\\
PKS 1424-41                 	&	4FGL J1427.9-4206	&	236699 	&	2.17 	$\pm$	0.01 	&	1.55 	&	2.15 	$\pm$	0.01 	&	0.07 	$\pm$	0.00 	&	5.07 	$\pm$	0.03 	&	48.47 	\\
4C +01.02                   	&	4FGL J0108.6+0134	&	106556 	&	14.27 	$\pm$	0.10 	&	0.46 	&	2.27 	$\pm$	0.01 	&	0.08 	$\pm$	0.01 	&	3.33 	$\pm$	0.05 	&	48.46 	\\
PKS 1502+106                	&	4FGL J1504.4+1029	&	88006 	&	8.85 	$\pm$	0.07 	&	0.53 	&	2.16 	$\pm$	0.01 	&	0.07 	$\pm$	0.01 	&	2.58 	$\pm$	0.02 	&	48.28 	\\
4C +38.41                   	&	4FGL J1635.2+3808	&	78168 	&	5.22 	$\pm$	0.04 	&	0.61 	&	2.37 	$\pm$	0.01 	&	0.09 	$\pm$	0.01 	&	2.40 	$\pm$	0.03 	&	48.16 	\\
PKS 0454-234                	&	4FGL J0457.0-2324	&	107156 	&	5.10 	$\pm$	0.04 	&	0.69 	&	2.11 	$\pm$	0.01 	&	0.08 	$\pm$	0.01 	&	2.36 	$\pm$	0.03 	&	47.70 	\\
4C +21.35                   	&	4FGL J1224.9+2122	&	67198 	&	10.34 	$\pm$	0.08 	&	0.44 	&	2.28 	$\pm$	0.01 	&	0.03 	$\pm$	0.01 	&	2.24 	$\pm$	0.02 	&	46.79 	\\
3C 273                      	&	4FGL J1229.0+0202	&	26629 	&	11.98 	$\pm$	0.13 	&	0.37 	&	2.63 	$\pm$	0.01 	&	0.07 	$\pm$	0.01 	&	2.17 	$\pm$	0.02 	&	45.72 	\\
Ton 599                     	&	4FGL J1159.5+2914	&	103958 	&	6.10 	$\pm$	0.05 	&	0.61 	&	2.03 	$\pm$	0.01 	&	0.07 	$\pm$	0.01 	&	2.07 	$\pm$	0.02 	&	47.37 	\\
PKS 0402-362                	&	4FGL J0403.9-3605	&	48185 	&	11.18 	$\pm$	0.10 	&	0.38 	&	2.41 	$\pm$	0.01 	&	0.10 	$\pm$	0.01 	&	1.83 	$\pm$	0.02 	&	47.75 	\\
B2 1520+31                  	&	4FGL J1522.1+3144	&	51648 	&	5.05 	$\pm$	0.05 	&	0.53 	&	2.36 	$\pm$	0.01 	&	0.05 	$\pm$	0.01 	&	1.79 	$\pm$	0.02 	&	47.83 	\\
4C +28.07                   	&	4FGL J0237.8+2848	&	43942 	&	5.82 	$\pm$	0.06 	&	0.54 	&	2.19 	$\pm$	0.01 	&	0.08 	$\pm$	0.01 	&	1.78 	$\pm$	0.03 	&	47.69 	\\
PKS 0346-27                 	&	4FGL J0348.6-2749	&	98722 	&	4.84 	$\pm$	0.04 	&	0.64 	&	1.96 	$\pm$	0.01 	&	0.06 	$\pm$	0.01 	&	1.72 	$\pm$	0.02 	&	47.62 	\\
S5 1044+71                  	&	4FGL J1048.4+7143	&	65615 	&	2.52 	$\pm$	0.02 	&	0.76 	&	2.20 	$\pm$	0.01 	&	0.07 	$\pm$	0.01 	&	1.55 	$\pm$	0.02 	&	47.61 	\\
PKS 0208-512                	&	4FGL J0210.7-5101	&	50939 	&	4.67 	$\pm$	0.05 	&	0.55 	&	2.22 	$\pm$	0.01 	&	0.08 	$\pm$	0.01 	&	1.50 	$\pm$	0.02 	&	47.43 	\\
PKS 2326-502                	&	4FGL J2329.3-4955	&	43616 	&	4.22 	$\pm$	0.04 	&	0.56 	&	2.19 	$\pm$	0.01 	&	0.09 	$\pm$	0.01 	&	1.37 	$\pm$	0.02 	&	46.78 	\\
PKS 2023-07                 	&	4FGL J2025.6-0735	&	26062 	&	2.80 	$\pm$	0.04 	&	0.66 	&	2.20 	$\pm$	0.01 	&	0.07 	$\pm$	0.01 	&	1.31 	$\pm$	0.02 	&	47.71 	\\
PKS 2052-47                 	&	4FGL J2056.2-4714	&	21583 	&	5.01 	$\pm$	0.06 	&	0.48 	&	2.32 	$\pm$	0.01 	&	0.12 	$\pm$	0.01 	&	1.23 	$\pm$	0.02 	&	47.67 	\\
B3 1343+451                 	&	4FGL J1345.5+4453	&	41868 	&	4.14 	$\pm$	0.04 	&	0.51 	&	2.19 	$\pm$	0.01 	&	0.04 	$\pm$	0.01 	&	1.19 	$\pm$	0.02 	&	48.25 	\\
PKS 0805-07                 	&	4FGL J0808.2-0751	&	24790 	&	2.93 	$\pm$	0.04 	&	0.64 	&	2.11 	$\pm$	0.01 	&	0.06 	$\pm$	0.01 	&	1.19 	$\pm$	0.03 	&	47.98 	\\
S3 0458-02                  	&	4FGL J0501.2-0158	&	18705 	&	4.62 	$\pm$	0.06 	&	0.47 	&	2.30 	$\pm$	0.01 	&	0.05 	$\pm$	0.01 	&	1.18 	$\pm$	0.02 	&	48.09 	\\
PKS 0736+01                 	&	4FGL J0739.2+0137	&	18420 	&	3.13 	$\pm$	0.04 	&	0.60 	&	2.25 	$\pm$	0.02 	&	0.13 	$\pm$	0.01 	&	1.16 	$\pm$	0.03 	&	45.76 	\\
PKS 1244-255                	&	4FGL J1246.7-2548	&	20590 	&	2.17 	$\pm$	0.03 	&	0.69 	&	2.22 	$\pm$	0.01 	&	0.09 	$\pm$	0.01 	&	1.09 	$\pm$	0.02 	&	46.88 	\\
B2 0218+357                 	&	4FGL J0221.1+3556	&	24168 	&	1.40 	$\pm$	0.02 	&	0.82 	&	2.21 	$\pm$	0.01 	&	0.07 	$\pm$	0.01 	&	1.04 	$\pm$	0.03 	&	47.25 	\\
4C +71.07                   	&	4FGL J0841.3+7053	&	14003 	&	4.60 	$\pm$	0.07 	&	0.39 	&	2.81 	$\pm$	0.02 	&	0.15 	$\pm$	0.02 	&	1.01 	$\pm$	0.02 	&	47.82 	\\
PKS 1824-582                	&	4FGL J1829.2-5813	&	8000 	&	1.30 	$\pm$	0.03 	&	0.66 	&	2.56 	$\pm$	0.02 	&	0.04 	$\pm$	0.01 	&	0.98 	$\pm$	0.02 	&	47.53 	\\
PKS 0502+049                	&	4FGL J0505.3+0459	&	10716 	&	0.97 	$\pm$	0.02 	&	0.86 	&	2.32 	$\pm$	0.01 	&	0.05 	$\pm$	0.01 	&	0.97 	$\pm$	0.04 	&	47.17 	\\
PKS 0336-01                 	&	4FGL J0339.5-0146	&	14801 	&	3.34 	$\pm$	0.05 	&	0.51 	&	2.25 	$\pm$	0.01 	&	0.04 	$\pm$	0.01 	&	0.97 	$\pm$	0.02 	&	47.07 	\\
PKS 2227-08                 	&	4FGL J2229.7-0832	&	7709 	&	6.22 	$\pm$	0.12 	&	0.35 	&	2.60 	$\pm$	0.02 	&	0.07 	$\pm$	0.02 	&	0.96 	$\pm$	0.02 	&	47.51 	\\
PKS 1954-388                	&	4FGL J1958.0-3845	&	17643 	&	2.04 	$\pm$	0.03 	&	0.70 	&	2.14 	$\pm$	0.01 	&	0.10 	$\pm$	0.01 	&	0.95 	$\pm$	0.02 	&	46.85 	\\
OX 169                      	&	4FGL J2143.5+1743	&	9135 	&	3.65 	$\pm$	0.07 	&	0.45 	&	2.45 	$\pm$	0.02 	&	0.03 	$\pm$	0.01 	&	0.95 	$\pm$	0.02 	&	45.69 	\\
OG 050                      	&	4FGL J0532.6+0732	&	7592 	&	1.94 	$\pm$	0.04 	&	0.63 	&	2.28 	$\pm$	0.02 	&	0.06 	$\pm$	0.01 	&	0.88 	$\pm$	0.03 	&	47.40 	\\
S4 1144+40                  	&	4FGL J1146.9+3958	&	17979 	&	2.27 	$\pm$	0.03 	&	0.56 	&	2.34 	$\pm$	0.01 	&	0.05 	$\pm$	0.01 	&	0.88 	$\pm$	0.02 	&	47.23 	\\
OQ 334                      	&	4FGL J1422.5+3223	&	27130 	&	2.65 	$\pm$	0.03 	&	0.56 	&	2.12 	$\pm$	0.01 	&	0.06 	$\pm$	0.01 	&	0.84 	$\pm$	0.01 	&	46.86 	\\
4C +31.03                   	&	4FGL J0112.8+3208	&	18632 	&	2.52 	$\pm$	0.04 	&	0.56 	&	2.09 	$\pm$	0.02 	&	0.05 	$\pm$	0.01 	&	0.78 	$\pm$	0.02 	&	46.73 	\\
PKS 1127-14                 	&	4FGL J1129.8-1447	&	6626 	&	1.27 	$\pm$	0.03 	&	0.63 	&	2.56 	$\pm$	0.02 	&	0.09 	$\pm$	0.02 	&	0.76 	$\pm$	0.06 	&	47.18 	\\
MG1 J123931+0443            	&	4FGL J1239.5+0443	&	12707 	&	2.76 	$\pm$	0.05 	&	0.53 	&	2.17 	$\pm$	0.02 	&	0.12 	$\pm$	0.01 	&	0.76 	$\pm$	0.02 	&	47.70 	\\
PKS B1406-076               	&	4FGL J1408.9-0751	&	19737 	&	0.96 	$\pm$	0.02 	&	0.99 	&	2.13 	$\pm$	0.01 	&	0.16 	$\pm$	0.01 	&	0.75 	$\pm$	0.02 	&	47.63 	\\
4C +55.17                   	&	4FGL J0957.6+5523	&	45819 	&	1.51 	$\pm$	0.02 	&	0.81 	&	1.86 	$\pm$	0.01 	&	0.07 	$\pm$	0.01 	&	0.74 	$\pm$	0.01 	&	47.23 	\\
PKS 0440-00                 	&	4FGL J0442.6-0017	&	7692 	&	3.13 	$\pm$	0.06 	&	0.44 	&	2.39 	$\pm$	0.02 	&	0.03 	$\pm$	0.01 	&	0.74 	$\pm$	0.02 	&	46.29 	\\
PKS B1908-201               	&	4FGL J1911.2-2006	&	5686 	&	1.27 	$\pm$	0.03 	&	0.71 	&	2.33 	$\pm$	0.02 	&	0.12 	$\pm$	0.02 	&	0.73 	$\pm$	0.03 	&	47.20 	\\
PKS 2142-75                 	&	4FGL J2147.3-7536	&	7842 	&	0.71 	$\pm$	0.02 	&	0.81 	&	2.39 	$\pm$	0.02 	&	0.02 	$\pm$	0.01 	&	0.73 	$\pm$	0.05 	&	47.18 	\\
3C 345                      	&	4FGL J1642.9+3948	&	9784 	&	0.30 	$\pm$	0.01 	&	1.25 	&	2.37 	$\pm$	0.02 	&	0.05 	$\pm$	0.01 	&	0.72 	$\pm$	0.05 	&	46.59 	\\
PMN J2345-1555              	&	4FGL J2345.2-1555	&	18420 	&	1.05 	$\pm$	0.02 	&	0.91 	&	2.01 	$\pm$	0.01 	&	0.09 	$\pm$	0.01 	&	0.71 	$\pm$	0.03 	&	46.79 	\\
PKS 2345-16                 	&	4FGL J2348.0-1630	&	11260 	&	0.88 	$\pm$	0.02 	&	0.87 	&	2.18 	$\pm$	0.01 	&	0.07 	$\pm$	0.01 	&	0.71 	$\pm$	0.04 	&	46.63 	\\
PKS 2032+107                	&	4FGL J2035.4+1056	&	6402 	&	0.90 	$\pm$	0.02 	&	0.78 	&	2.35 	$\pm$	0.02 	&	0.08 	$\pm$	0.02 	&	0.70 	$\pm$	0.03 	&	46.59 	\\
B2 2234+28A                 	&	4FGL J2236.3+2828	&	16006 	&	1.71 	$\pm$	0.03 	&	0.66 	&	2.08 	$\pm$	0.01 	&	0.10 	$\pm$	0.01 	&	0.67 	$\pm$	0.02 	&	46.93 	\\
TXS 1700+685                	&	4FGL J1700.0+6830	&	18464 	&	1.22 	$\pm$	0.02 	&	0.70 	&	2.22 	$\pm$	0.01 	&	0.06 	$\pm$	0.01 	&	0.67 	$\pm$	0.02 	&	45.97 	\\
PKS 0438-43                 	&	4FGL J0440.3-4332	&	7738 	&	1.08 	$\pm$	0.02 	&	0.65 	&	2.57 	$\pm$	0.02 	&	0.13 	$\pm$	0.02 	&	0.66 	$\pm$	0.03 	&	47.98 	\\

\end{tabular}}
\end{adjustwidth}
\end{center}
\end{table}

\renewcommand\thetable{2}   
\begin{table}
\setlength{\abovecaptionskip}{0 cm}
\setlength{\belowcaptionskip}{0.1cm}
\begin{center}
\caption{Continued}
\centering
\small
\setlength{\LTleft}{0 cm} \setlength{\LTright}{0 cm} 
\begin{adjustwidth}{-2.2cm}{-5cm}
\resizebox{20cm}{!}{
\begin{tabular}{cccccccccc} \hline
\renewcommand\arraystretch{2}
\centering
{Source Name}&FGL Name & TS& $N_0$ & $E_0$ & $\alpha$ & $\beta$  &$\rm F_\gamma$&$\rm log_{10}(L_\gamma)$\\
\normalsize(1) & \normalsize(2) & \normalsize(3) &\normalsize(4) & \normalsize(5) &  \normalsize(6) &  \normalsize(7)    &  \normalsize(8)   &  \normalsize(9)   \\
\hline 
PKS 0308-611                	&	4FGL J0309.9-6058	&	9798 	&	1.46 	$\pm$	0.03 	&	0.58 	&	2.40 	$\pm$	0.02 	&	0.05 	$\pm$	0.01 	&	0.66 	$\pm$	0.02 	&	47.37 	\\
PKS 0250-225                	&	4FGL J0252.8-2219	&	10852 	&	3.10 	$\pm$	0.05 	&	0.44 	&	2.25 	$\pm$	0.02 	&	0.08 	$\pm$	0.01 	&	0.63 	$\pm$	0.01 	&	47.36 	\\
PKS 2255-282                	&	4FGL J2258.1-2759	&	11005 	&	0.91 	$\pm$	0.02 	&	0.87 	&	2.33 	$\pm$	0.02 	&	0.22 	$\pm$	0.02 	&	0.63 	$\pm$	0.03 	&	47.01 	\\
B2 2308+34                  	&	4FGL J2311.0+3425	&	9646 	&	1.79 	$\pm$	0.03 	&	0.57 	&	2.22 	$\pm$	0.02 	&	0.07 	$\pm$	0.01 	&	0.62 	$\pm$	0.02 	&	47.63 	\\
S4 1726+45                  	&	4FGL J1727.4+4530	&	10668 	&	2.08 	$\pm$	0.04 	&	0.50 	&	2.33 	$\pm$	0.02 	&	0.06 	$\pm$	0.01 	&	0.62 	$\pm$	0.03 	&	46.68 	\\
PKS 1124-186                	&	4FGL J1127.0-1857	&	11677 	&	1.58 	$\pm$	0.03 	&	0.63 	&	2.10 	$\pm$	0.02 	&	0.07 	$\pm$	0.01 	&	0.62 	$\pm$	0.02 	&	47.15 	\\
TXS 0025+197                	&	4FGL J0028.5+2001	&	11857 	&	1.58 	$\pm$	0.03 	&	0.64 	&	2.09 	$\pm$	0.02 	&	0.09 	$\pm$	0.01 	&	0.60 	$\pm$	0.02 	&	47.52 	\\
PKS 1454-354                	&	4FGL J1457.4-3539	&	7549 	&	1.21 	$\pm$	0.03 	&	0.71 	&	2.19 	$\pm$	0.02 	&	0.12 	$\pm$	0.02 	&	0.59 	$\pm$	0.03 	&	47.40 	\\
PKS 0202-17                 	&	4FGL J0205.0-1700	&	5784 	&	1.76 	$\pm$	0.04 	&	0.48 	&	2.48 	$\pm$	0.02 	&	0.03 	$\pm$	0.02 	&	0.56 	$\pm$	0.02 	&	47.43 	\\
S4 1030+61                  	&	4FGL J1033.9+6050	&	17228 	&	0.35 	$\pm$	0.01 	&	1.25 	&	2.16 	$\pm$	0.02 	&	0.06 	$\pm$	0.01 	&	0.56 	$\pm$	0.06 	&	47.41 	\\
S4 0917+44                  	&	4FGL J0920.9+4441	&	14991 	&	0.38 	$\pm$	0.01 	&	1.21 	&	2.37 	$\pm$	0.02 	&	0.19 	$\pm$	0.02 	&	0.55 	$\pm$	0.05 	&	47.81 	\\
OP 313                      	&	4FGL J1310.5+3221	&	15880 	&	0.28 	$\pm$	0.00 	&	1.61 	&	2.15 	$\pm$	0.02 	&	0.14 	$\pm$	0.02 	&	0.55 	$\pm$	0.04 	&	47.15 	\\
PKS 1004-217                	&	4FGL J1006.7-2159	&	5931 	&	1.05 	$\pm$	0.03 	&	0.63 	&	2.32 	$\pm$	0.02 	&	0.02 	$\pm$	0.02 	&	0.54 	$\pm$	0.05 	&	45.90 	\\
OK 630                      	&	4FGL J0921.6+6216	&	11782 	&	1.32 	$\pm$	0.02 	&	0.61 	&	2.18 	$\pm$	0.02 	&	0.06 	$\pm$	0.01 	&	0.51 	$\pm$	0.01 	&	47.34 	\\
PKS 0226-559                	&	4FGL J0228.3-5547	&	12260 	&	0.34 	$\pm$	0.01 	&	1.18 	&	2.21 	$\pm$	0.02 	&	0.09 	$\pm$	0.01 	&	0.51 	$\pm$	0.01 	&	47.90 	\\
PKS 0035-252                	&	4FGL J0038.2-2459	&	10290 	&	1.11 	$\pm$	0.02 	&	0.65 	&	2.16 	$\pm$	0.02 	&	0.05 	$\pm$	0.01 	&	0.50 	$\pm$	0.02 	&	47.16 	\\
PKS 0420-01                 	&	4FGL J0423.3-0120	&	6392 	&	1.65 	$\pm$	0.04 	&	0.58 	&	2.14 	$\pm$	0.03 	&	0.15 	$\pm$	0.02 	&	0.50 	$\pm$	0.02 	&	46.90 	\\
PKS 0524-485                	&	4FGL J0526.2-4830	&	9795 	&	0.63 	$\pm$	0.01 	&	0.86 	&	2.15 	$\pm$	0.02 	&	0.06 	$\pm$	0.01 	&	0.49 	$\pm$	0.02 	&	47.25 	\\
PKS 0215+015                	&	4FGL J0217.8+0144	&	8648 	&	0.94 	$\pm$	0.02 	&	0.70 	&	2.10 	$\pm$	0.02 	&	0.02 	$\pm$	0.01 	&	0.48 	$\pm$	0.02 	&	47.53 	\\
TXS 2241+406                	&	4FGL J2244.2+4057	&	8819 	&	0.34 	$\pm$	0.01 	&	1.25 	&	2.05 	$\pm$	0.02 	&	0.06 	$\pm$	0.01 	&	0.47 	$\pm$	0.02 	&	47.21 	\\
B2 2113+29                  	&	4FGL J2115.4+2932	&	5770 	&	0.71 	$\pm$	0.02 	&	0.83 	&	2.23 	$\pm$	0.02 	&	0.15 	$\pm$	0.02 	&	0.47 	$\pm$	0.01 	&	47.37 	\\
PKS 2320-035                	&	4FGL J2323.5-0317	&	7629 	&	1.19 	$\pm$	0.03 	&	0.66 	&	2.11 	$\pm$	0.02 	&	0.13 	$\pm$	0.02 	&	0.47 	$\pm$	0.02 	&	47.31 	\\
PKS 0116-219                	&	4FGL J0118.9-2141	&	5225 	&	0.26 	$\pm$	0.01 	&	1.05 	&	2.30 	$\pm$	0.02 	&	-0.02	$\pm$	0.02 	&	0.47 	$\pm$	0.02 	&	47.03 	\\
B2 0716+33                  	&	4FGL J0719.3+3307	&	8883 	&	0.81 	$\pm$	0.02 	&	0.77 	&	2.06 	$\pm$	0.02 	&	0.05 	$\pm$	0.01 	&	0.45 	$\pm$	0.02 	&	46.77 	\\
PKS 0405-385                	&	4FGL J0407.0-3826	&	8019 	&	1.45 	$\pm$	0.03 	&	0.55 	&	2.19 	$\pm$	0.02 	&	0.08 	$\pm$	0.02 	&	0.45 	$\pm$	0.02 	&	47.16 	\\
TXS 1318+225                	&	4FGL J1321.1+2216	&	5272 	&	0.64 	$\pm$	0.02 	&	0.71 	&	2.31 	$\pm$	0.02 	&	0.01 	$\pm$	0.02 	&	0.44 	$\pm$	0.02 	&	46.81 	\\
OC 457                      	&	4FGL J0137.0+4751	&	6021 	&	0.79 	$\pm$	0.02 	&	0.74 	&	2.14 	$\pm$	0.02 	&	0.09 	$\pm$	0.02 	&	0.42 	$\pm$	0.02 	&	46.79 	\\
4C +14.23                   	&	4FGL J0725.2+1425	&	5355 	&	1.27 	$\pm$	0.03 	&	0.60 	&	2.06 	$\pm$	0.03 	&	0.11 	$\pm$	0.02 	&	0.40 	$\pm$	0.02 	&	46.97 	\\
PKS 0131-522                	&	4FGL J0133.1-5201	&	7188 	&	0.90 	$\pm$	0.02 	&	0.64 	&	2.15 	$\pm$	0.02 	&	0.09 	$\pm$	0.02 	&	0.36 	$\pm$	0.02 	&	46.78 	\\
PKS 1441+25                 	&	4FGL J1443.9+2501	&	8677 	&	0.32 	$\pm$	0.01 	&	1.17 	&	2.02 	$\pm$	0.02 	&	0.08 	$\pm$	0.02 	&	0.35 	$\pm$	0.02 	&	46.90 	\\
PKS 0514-459                	&	4FGL J0515.6-4556	&	6415 	&	0.38 	$\pm$	0.01 	&	0.96 	&	2.19 	$\pm$	0.03 	&	0.09 	$\pm$	0.02 	&	0.35 	$\pm$	0.03 	&	45.33 	\\
B2 1732+38A                 	&	4FGL J1734.3+3858	&	6714 	&	0.65 	$\pm$	0.02 	&	0.78 	&	2.17 	$\pm$	0.02 	&	0.14 	$\pm$	0.02 	&	0.35 	$\pm$	0.02 	&	46.84 	\\
PKS 0244-470                	&	4FGL J0245.9-4650	&	5177 	&	1.09 	$\pm$	0.03 	&	0.56 	&	2.26 	$\pm$	0.03 	&	0.14 	$\pm$	0.02 	&	0.35 	$\pm$	0.01 	&	47.09 	\\
S4 1849+67                  	&	4FGL J1849.2+6705	&	7435 	&	0.34 	$\pm$	0.01 	&	1.07 	&	2.15 	$\pm$	0.02 	&	0.19 	$\pm$	0.02 	&	0.29 	$\pm$	0.01 	&	46.44 	\\
S5 1217+71                  	&	4FGL J1220.1+7105	&	5910 	&	0.51 	$\pm$	0.01 	&	0.75 	&	2.10 	$\pm$	0.03 	&	0.22 	$\pm$	0.02 	&	0.22 	$\pm$	0.01 	&	45.93 	\\

\hline
\multicolumn{9}{c}{Average}\\
\hline
Mean. / std.&& & 2.83 / 2.85 &0.72 / 0.25 & 2.24 / 0.16 & 0.08 / 0.04 & 1.28 / 1.70 &47.28\\
\hline
\end{tabular}}
\end{adjustwidth}
\end{center}
{\footnotesize{{Note. 
 \\
Column (4)is in unit of $\rm 10^{-11}\ cm^{-2}s^{-1}MeV^{-1}$; \\
Column (5) is in unit of $\rm GeV$; \\
Column (8) is in unit of $\rm cm^{-2}s^{-1}$;\\
Column (9) is in unit of $\rm erg\ s^{-1}$.\\
}}}
\end{table}

These 87 sources are the brightest FSRQs in the gamma-ray sky. The integrated photon flux in the 0.1-10 GeV range for these 87 bright FSRQ samples is larger than $ 2.16 \times 10^{-8}$ $\rm cm^{-2} s^{-1}$, with an average of $ \langle F_\gamma \rangle =(1.28 \pm 1.70) \times 10^{-7} \rm cm^{-2} s^{-1}$. The average gamma-ray luminosity is $\langle L_\gamma \rangle =(3.74 \pm 5.52) \times 10^{47} \rm erg\ s^{-1}$.

\subsection{BPL model fitting}

Using the Fermi spectral analysis pipeline described in Section 2.1, we also fit the gamma-ray spectra of these 87 bright sources with the BPL model:
\begin{equation}\label{eq4}
\frac{d N}{d E}=N_0 \times \begin{cases}\left(E / E_b\right)^{-\gamma_1} & \text { if } E<E_b \\ \left(E / E_b\right)^{-\gamma_2} & \text { otherwise }\end{cases}  \text {. }
\end{equation}
To locate the energy of the GeV spectral break, the low energy spectrum index ($\gamma_1$ of Equation \ref{eq4}), the high energy spectrum index ($\gamma_2$ of Equation \ref{eq4}), the spectral break energy ($E_b$ of Equation \ref{eq4}), and the differential flux at $E_b$ ($ N_0$ of Equation \ref{eq4}) were all freed during spectral fitting with the BPL model. The complete BPL fitting results for the 87 bright FSRQs are tabulated in Table \ref{tab:3}. The change in the photon index ($\Delta \gamma = \gamma_2 - \gamma_1$) and the break energy ($E_b^{\prime}=(1+z)E_b$) in the Cosmological rest frame are also listed in Table \ref{tab:3}.

In Figure \ref{fig:fig1}, the BPL best-fitting lines in the $E- E^2 dN/dE$ frame for these 87 sources are represented with green lines, and the corresponding 1-$\sigma$ confidence regions are shown in green. We extend the low-energy power-law component in the BPL model with a black dashed line to clearly demonstrate the break phenomenon of the gamma-ray spectrum. The energy band of the spectral break is marked by the gray area.

Additionally, we provide the actual SED data points of these FSRQs in the energy range from 100 MeV to 10 GeV using the \emph{sed()} tool in \emph{Fermipy}. The energy range is divided into 12 equally spaced energy bins, and we evaluated the TS of each source in each energy bin. For energy bins with a confidence level greater than 5 $\sigma$ ($\rm TS \geq 25$), we provided the SED data points, displayed as black points; for energy bins with a confidence level greater than 3 $\sigma$ ($\rm 9 \leq TS < 25$), we provided the SED data points, displayed as blue points; for energy bins with a confidence level less than 3 $\sigma$ ($\rm 0 < TS < 9$), we provided the 95$\%$ confidence level upper limit of the SED, displayed as black arrows. Energy bins where no photons were detected were excluded from the analysis. All actual SED data points are shown in Figure \ref{fig:fig1}.

\renewcommand\thetable{3}
\begin{table}
\setlength{\abovecaptionskip}{0 cm}
\setlength{\belowcaptionskip}{0.1cm}
\begin{center}
\caption{BPL fitting  results}
\label{tab:3}
\centering
\small
\setlength{\LTleft}{0 cm} \setlength{\LTright}{0 cm} 
\begin{adjustwidth}{-2.2cm}{-5cm}
\resizebox{20cm}{!}{
\begin{tabular}{ccccccccccc} \hline
\renewcommand\arraystretch{2}
\centering
{Source Name}&Redshift& $N_0$  & $\gamma_1$ &  $\gamma_2$  &$\Delta \gamma$&$\rm E_b$&$\rm E_b^{\prime}$&$\rm \Delta AIC$\\
\normalsize(1) & \normalsize(2) & \normalsize(3) &\normalsize(4) & \normalsize(5) &  \normalsize(6) &  \normalsize(7)    &  \normalsize(8)   &  \normalsize(9)   \\

\hline 
3C 454.3                    	&	0.859 	&	12.58 	$\pm$	0.82 	&	2.19 	$\pm$	0.01 	&	2.62 	$\pm$	0.01 	&	0.43 	$\pm$	0.01 	&	0.91 	$\pm$	0.03 	&	1.69 	$\pm$	0.05 	&	-1305.35 	\\
PKS 1510-089                	&	0.360 	&	2.56 	$\pm$	0.10 	&	2.31 	$\pm$	0.01 	&	2.50 	$\pm$	0.02 	&	0.19 	$\pm$	0.01 	&	1.25 	$\pm$	0.02 	&	1.70 	$\pm$	0.03 	&	117.58 	\\
3C 279                      	&	0.536 	&	6.34 	$\pm$	0.74 	&	2.13 	$\pm$	0.01 	&	2.43 	$\pm$	0.01 	&	0.30 	$\pm$	0.01 	&	0.94 	$\pm$	0.05 	&	1.44 	$\pm$	0.07 	&	-2033.15 	\\
CTA 102                     	&	1.037 	&	10.44 	$\pm$	1.50 	&	2.13 	$\pm$	0.01 	&	2.41 	$\pm$	0.01 	&	0.28 	$\pm$	0.01 	&	0.74 	$\pm$	0.05 	&	1.50 	$\pm$	0.10 	&	-1529.33 	\\
PKS 1424-41                 	&	1.522 	&	10.13 	$\pm$	1.15 	&	1.92 	$\pm$	0.01 	&	2.21 	$\pm$	0.01 	&	0.28 	$\pm$	0.01 	&	0.76 	$\pm$	0.04 	&	1.92 	$\pm$	0.11 	&	-873.91 	\\
4C +01.02                   	&	2.099 	&	2.30 	$\pm$	0.37 	&	2.22 	$\pm$	0.01 	&	2.60 	$\pm$	0.03 	&	0.37 	$\pm$	0.02 	&	1.04 	$\pm$	0.07 	&	3.23 	$\pm$	0.23 	&	-7.50 	\\
PKS 1502+106                	&	1.839 	&	2.73 	$\pm$	0.52 	&	2.07 	$\pm$	0.01 	&	2.40 	$\pm$	0.02 	&	0.33 	$\pm$	0.02 	&	0.94 	$\pm$	0.08 	&	2.67 	$\pm$	0.23 	&	-39.17 	\\
4C +38.41                   	&	1.814 	&	1.29 	$\pm$	0.21 	&	2.27 	$\pm$	0.01 	&	2.73 	$\pm$	0.03 	&	0.45 	$\pm$	0.02 	&	1.13 	$\pm$	0.08 	&	3.17 	$\pm$	0.22 	&	160.68 	\\
PKS 0454-234                	&	1.003 	&	8.12 	$\pm$	1.26 	&	1.92 	$\pm$	0.02 	&	2.26 	$\pm$	0.01 	&	0.34 	$\pm$	0.02 	&	0.57 	$\pm$	0.04 	&	1.14 	$\pm$	0.08 	&	-4.54 	\\
4C +21.35                   	&	0.434 	&	0.28 	$\pm$	0.14 	&	2.26 	$\pm$	0.01 	&	2.60 	$\pm$	0.08 	&	0.33 	$\pm$	0.03 	&	2.12 	$\pm$	0.44 	&	3.05 	$\pm$	0.64 	&	32.29 	\\
3C 273                      	&	0.158 	&	0.67 	$\pm$	0.30 	&	2.64 	$\pm$	0.01 	&	3.04 	$\pm$	0.07 	&	0.40 	$\pm$	0.03 	&	1.10 	$\pm$	0.18 	&	1.28 	$\pm$	0.21 	&	314.95 	\\
Ton 599                     	&	0.729 	&	2.17 	$\pm$	0.39 	&	1.95 	$\pm$	0.01 	&	2.24 	$\pm$	0.02 	&	0.28 	$\pm$	0.02 	&	1.03 	$\pm$	0.09 	&	1.87 	$\pm$	0.15 	&	-2.38 	\\
PKS 0402-362                	&	1.417 	&	0.98 	$\pm$	0.18 	&	2.37 	$\pm$	0.01 	&	2.97 	$\pm$	0.05 	&	0.60 	$\pm$	0.02 	&	1.05 	$\pm$	0.08 	&	2.54 	$\pm$	0.19 	&	220.06 	\\
B2 1520+31                  	&	1.489 	&	0.33 	$\pm$	0.08 	&	2.32 	$\pm$	0.01 	&	2.77 	$\pm$	0.06 	&	0.45 	$\pm$	0.02 	&	1.72 	$\pm$	0.18 	&	4.29 	$\pm$	0.45 	&	16.52 	\\
4C +28.07                   	&	1.206 	&	2.49 	$\pm$	0.57 	&	2.11 	$\pm$	0.02 	&	2.42 	$\pm$	0.02 	&	0.31 	$\pm$	0.02 	&	0.81 	$\pm$	0.08 	&	1.79 	$\pm$	0.18 	&	50.39 	\\
PKS 0346-27                 	&	0.991 	&	2.38 	$\pm$	0.56 	&	1.87 	$\pm$	0.02 	&	2.15 	$\pm$	0.02 	&	0.28 	$\pm$	0.02 	&	0.94 	$\pm$	0.11 	&	1.87 	$\pm$	0.22 	&	-9.58 	\\
S5 1044+71                  	&	1.150 	&	3.58 	$\pm$	0.24 	&	2.04 	$\pm$	0.02 	&	2.34 	$\pm$	0.01 	&	0.30 	$\pm$	0.02 	&	0.66 	$\pm$	0.02 	&	1.41 	$\pm$	0.04 	&	9.47 	\\
PKS 0208-512                	&	1.003 	&	0.60 	$\pm$	0.31 	&	2.16 	$\pm$	0.02 	&	2.58 	$\pm$	0.08 	&	0.41 	$\pm$	0.04 	&	1.40 	$\pm$	0.32 	&	2.80 	$\pm$	0.64 	&	24.99 	\\
PKS 2326-502                	&	0.518 	&	1.09 	$\pm$	0.25 	&	2.10 	$\pm$	0.02 	&	2.52 	$\pm$	0.04 	&	0.42 	$\pm$	0.02 	&	1.06 	$\pm$	0.11 	&	1.61 	$\pm$	0.17 	&	-29.52 	\\
PKS 2023-07                 	&	1.388 	&	0.89 	$\pm$	0.23 	&	2.10 	$\pm$	0.02 	&	2.47 	$\pm$	0.04 	&	0.37 	$\pm$	0.03 	&	1.14 	$\pm$	0.13 	&	2.72 	$\pm$	0.31 	&	-10.59 	\\
PKS 2052-47                 	&	1.489 	&	2.48 	$\pm$	0.55 	&	2.17 	$\pm$	0.03 	&	2.66 	$\pm$	0.03 	&	0.49 	$\pm$	0.03 	&	0.66 	$\pm$	0.06 	&	1.65 	$\pm$	0.15 	&	23.46 	\\
B3 1343+451                 	&	2.534 	&	0.28 	$\pm$	0.08 	&	2.16 	$\pm$	0.01 	&	2.46 	$\pm$	0.05 	&	0.30 	$\pm$	0.02 	&	1.76 	$\pm$	0.22 	&	6.23 	$\pm$	0.79 	&	27.22 	\\
PKS 0805-07                 	&	1.837 	&	0.88 	$\pm$	0.31 	&	2.03 	$\pm$	0.02 	&	2.33 	$\pm$	0.04 	&	0.30 	$\pm$	0.03 	&	1.16 	$\pm$	0.19 	&	3.28 	$\pm$	0.53 	&	-6.92 	\\
S3 0458-02                  	&	2.291 	&	0.51 	$\pm$	0.28 	&	2.25 	$\pm$	0.02 	&	2.57 	$\pm$	0.07 	&	0.32 	$\pm$	0.04 	&	1.25 	$\pm$	0.30 	&	4.10 	$\pm$	0.98 	&	-11.38 	\\
PKS 0736+01                 	&	0.189 	&	4.08 	$\pm$	0.90 	&	1.98 	$\pm$	0.05 	&	2.51 	$\pm$	0.03 	&	0.53 	$\pm$	0.04 	&	0.55 	$\pm$	0.05 	&	0.66 	$\pm$	0.06 	&	35.96 	\\
PKS 1244-255                	&	0.635 	&	0.44 	$\pm$	0.14 	&	2.12 	$\pm$	0.02 	&	2.62 	$\pm$	0.07 	&	0.50 	$\pm$	0.04 	&	1.45 	$\pm$	0.20 	&	2.38 	$\pm$	0.32 	&	7.62 	\\
B2 0218+357                 	&	0.944 	&	5.33 	$\pm$	0.55 	&	1.84 	$\pm$	0.05 	&	2.33 	$\pm$	0.02 	&	0.49 	$\pm$	0.04 	&	0.46 	$\pm$	0.03 	&	0.90 	$\pm$	0.05 	&	-15.38 	\\
4C +71.07                   	&	2.218 	&	0.59 	$\pm$	0.14 	&	2.67 	$\pm$	0.02 	&	3.62 	$\pm$	0.11 	&	0.95 	$\pm$	0.05 	&	0.83 	$\pm$	0.07 	&	2.68 	$\pm$	0.21 	&	74.38 	\\
PKS 1824-582                	&	1.531 	&	0.32 	$\pm$	0.05 	&	2.47 	$\pm$	0.02 	&	2.87 	$\pm$	0.06 	&	0.31 	$\pm$	0.04 	&	1.18 	$\pm$	0.08 	&	2.99 	$\pm$	0.19 	&	41.11 	\\
PKS 0502+049                	&	0.954 	&	0.55 	$\pm$	0.42 	&	2.20 	$\pm$	0.04 	&	2.52 	$\pm$	0.07 	&	0.32 	$\pm$	0.05 	&	1.13 	$\pm$	0.37 	&	2.22 	$\pm$	0.72 	&	29.59 	\\
PKS 0336-01                 	&	0.850 	&	0.22 	$\pm$	0.17 	&	2.21 	$\pm$	0.02 	&	2.54 	$\pm$	0.10 	&	0.32 	$\pm$	0.05 	&	1.71 	$\pm$	0.56 	&	3.16 	$\pm$	1.04 	&	8.86 	\\
PKS 2227-08                 	&	1.560 	&	0.41 	$\pm$	0.07 	&	2.58 	$\pm$	0.02 	&	2.94 	$\pm$	0.07 	&	0.36 	$\pm$	0.04 	&	1.00 	$\pm$	0.06 	&	2.56 	$\pm$	0.16 	&	59.31 	\\
PKS 1954-388                	&	0.630 	&	0.60 	$\pm$	0.16 	&	2.02 	$\pm$	0.02 	&	2.51 	$\pm$	0.05 	&	0.49 	$\pm$	0.03 	&	1.27 	$\pm$	0.15 	&	2.06 	$\pm$	0.25 	&	-86.25 	\\
OX 169                      	&	0.211 	&	0.95 	$\pm$	1.44 	&	2.38 	$\pm$	0.05 	&	2.59 	$\pm$	0.08 	&	0.21 	$\pm$	0.06 	&	0.79 	$\pm$	0.49 	&	0.95 	$\pm$	0.59 	&	-34.97 	\\
OG 050                      	&	1.254 	&	1.31 	$\pm$	0.87 	&	2.15 	$\pm$	0.05 	&	2.45 	$\pm$	0.05 	&	0.30 	$\pm$	0.05 	&	0.76 	$\pm$	0.22 	&	1.72 	$\pm$	0.50 	&	22.30 	\\
S4 1144+40                  	&	1.089 	&	0.12 	$\pm$	0.08 	&	2.31 	$\pm$	0.02 	&	2.71 	$\pm$	0.14 	&	0.40 	$\pm$	0.05 	&	1.96 	$\pm$	0.57 	&	4.09 	$\pm$	1.18 	&	-10.08 	\\
OQ 334                      	&	0.682 	&	0.20 	$\pm$	0.11 	&	2.10 	$\pm$	0.02 	&	2.45 	$\pm$	0.09 	&	0.34 	$\pm$	0.04 	&	1.87 	$\pm$	0.47 	&	3.14 	$\pm$	0.79 	&	4.73 	\\
4C +31.03                   	&	0.603 	&	0.10 	$\pm$	0.09 	&	2.09 	$\pm$	0.02 	&	2.49 	$\pm$	0.21 	&	0.40 	$\pm$	0.06 	&	2.58 	$\pm$	1.08 	&	4.13 	$\pm$	1.74 	&	7.38 	\\
PKS 1127-14                 	&	1.184 	&	0.19 	$\pm$	0.03 	&	2.45 	$\pm$	0.02 	&	3.04 	$\pm$	0.10 	&	0.60 	$\pm$	0.05 	&	1.36 	$\pm$	0.09 	&	2.98 	$\pm$	0.20 	&	-5.61 	\\
MG1 J123931+0443            	&	1.761 	&	0.51 	$\pm$	0.13 	&	2.09 	$\pm$	0.02 	&	2.63 	$\pm$	0.06 	&	0.54 	$\pm$	0.04 	&	1.17 	$\pm$	0.13 	&	3.24 	$\pm$	0.37 	&	43.21 	\\
PKS B1406-076               	&	1.493 	&	0.96 	$\pm$	0.20 	&	1.88 	$\pm$	0.03 	&	2.44 	$\pm$	0.04 	&	0.56 	$\pm$	0.04 	&	1.03 	$\pm$	0.10 	&	2.56 	$\pm$	0.26 	&	18.55 	\\
4C +55.17                   	&	0.896 	&	0.82 	$\pm$	0.21 	&	1.77 	$\pm$	0.02 	&	2.07 	$\pm$	0.03 	&	0.30 	$\pm$	0.03 	&	1.15 	$\pm$	0.16 	&	2.18 	$\pm$	0.30 	&	-29.38 	\\
PKS 0440-00                 	&	0.449 	&	0.33 	$\pm$	0.24 	&	2.30 	$\pm$	0.04 	&	2.63 	$\pm$	0.08 	&	0.33 	$\pm$	0.06 	&	1.16 	$\pm$	0.35 	&	1.68 	$\pm$	0.51 	&	-14.43 	\\
PKS B1908-201               	&	1.119 	&	0.75 	$\pm$	0.25 	&	2.14 	$\pm$	0.04 	&	2.66 	$\pm$	0.06 	&	0.52 	$\pm$	0.05 	&	0.92 	$\pm$	0.13 	&	1.96 	$\pm$	0.28 	&	13.64 	\\
PKS 2142-75                 	&	1.138 	&	0.25 	$\pm$	0.24 	&	2.28 	$\pm$	0.03 	&	2.60 	$\pm$	0.11 	&	0.32 	$\pm$	0.06 	&	1.30 	$\pm$	0.51 	&	2.87 	$\pm$	1.08 	&	-56.83 	\\
3C 345                      	&	0.593 	&	0.38 	$\pm$	0.21 	&	2.19 	$\pm$	0.03 	&	2.55 	$\pm$	0.07 	&	0.35 	$\pm$	0.04 	&	1.17 	$\pm$	0.28 	&	1.87 	$\pm$	0.44 	&	-150.08 	\\
PMN J2345-1555              	&	0.621 	&	0.70 	$\pm$	0.19 	&	1.88 	$\pm$	0.03 	&	2.25 	$\pm$	0.04 	&	0.36 	$\pm$	0.03 	&	1.14 	$\pm$	0.15 	&	1.86 	$\pm$	0.25 	&	-17.71 	\\
PKS 2345-16                 	&	0.576 	&	0.61 	$\pm$	0.35 	&	2.07 	$\pm$	0.03 	&	2.37 	$\pm$	0.06 	&	0.30 	$\pm$	0.04 	&	1.05 	$\pm$	0.28 	&	1.66 	$\pm$	0.44 	&	-14.47 	\\
PKS 2032+107                	&	0.601 	&	0.11 	$\pm$	0.05 	&	2.25 	$\pm$	0.03 	&	2.89 	$\pm$	0.16 	&	0.64 	$\pm$	0.06 	&	2.01 	$\pm$	0.38 	&	3.21 	$\pm$	0.61 	&	25.22 	\\
B2 2234+28A                 	&	0.790 	&	0.36 	$\pm$	0.09 	&	1.99 	$\pm$	0.02 	&	2.49 	$\pm$	0.06 	&	0.50 	$\pm$	0.04 	&	1.42 	$\pm$	0.18 	&	2.55 	$\pm$	0.31 	&	14.74 	\\
TXS 1700+685                	&	0.301 	&	0.29 	$\pm$	0.13 	&	2.13 	$\pm$	0.02 	&	2.52 	$\pm$	0.07 	&	0.40 	$\pm$	0.04 	&	1.38 	$\pm$	0.28 	&	1.80 	$\pm$	0.36 	&	1.32 	\\
PKS 0438-43                 	&	2.863 	&	0.21 	$\pm$	0.06 	&	2.41 	$\pm$	0.03 	&	3.22 	$\pm$	0.13 	&	0.82 	$\pm$	0.06 	&	1.29 	$\pm$	0.15 	&	4.97 	$\pm$	0.58 	&	32.86 	\\

\end{tabular}}
\end{adjustwidth}
\end{center}
\end{table}

\renewcommand\thetable{3}
\begin{table}
\setlength{\abovecaptionskip}{0 cm}
\setlength{\belowcaptionskip}{0.1cm}
\begin{center}
\caption{Continued}
\centering
\small
\setlength{\LTleft}{0 cm} \setlength{\LTright}{0 cm} 
\begin{adjustwidth}{-2.2cm}{-5cm}
\resizebox{20cm}{!}{
\begin{tabular}{ccccccccccc} \hline
\renewcommand\arraystretch{2}
\centering
{Source Name}&Redshift& $N_0$  & $\gamma_1$ &  $\gamma_2$  &$\Delta \gamma$&$\rm E_b$&$\rm E_b^{\prime}$&$\rm \Delta AIC$\\
\normalsize(1) & \normalsize(2) & \normalsize(3) &\normalsize(4) & \normalsize(5) &  \normalsize(6) &  \normalsize(7)    &  \normalsize(8)   &  \normalsize(9)   \\

\hline 
PKS 0308-611                	&	1.479 	&	0.13 	$\pm$	0.02 	&	2.36 	$\pm$	0.02 	&	2.72 	$\pm$	0.08 	&	0.36 	$\pm$	0.04 	&	1.58 	$\pm$	0.10 	&	3.93 	$\pm$	0.26 	&	17.01 	\\
PKS 0250-225                	&	1.419 	&	0.36 	$\pm$	0.18 	&	2.20 	$\pm$	0.03 	&	2.63 	$\pm$	0.07 	&	0.43 	$\pm$	0.05 	&	1.14 	$\pm$	0.24 	&	2.76 	$\pm$	0.57 	&	-0.96 	\\
PKS 2255-282                	&	0.926 	&	0.15 	$\pm$	0.04 	&	2.24 	$\pm$	0.02 	&	2.98 	$\pm$	0.13 	&	0.75 	$\pm$	0.05 	&	1.87 	$\pm$	0.24 	&	3.60 	$\pm$	0.46 	&	32.55 	\\
B2 2308+34                  	&	1.817 	&	0.32 	$\pm$	0.28 	&	2.15 	$\pm$	0.04 	&	2.52 	$\pm$	0.11 	&	0.37 	$\pm$	0.06 	&	1.26 	$\pm$	0.49 	&	3.54 	$\pm$	1.37 	&	16.97 	\\
S4 1726+45                  	&	0.717 	&	0.07 	$\pm$	0.03 	&	2.29 	$\pm$	0.02 	&	2.92 	$\pm$	0.14 	&	0.63 	$\pm$	0.05 	&	2.13 	$\pm$	0.35 	&	3.66 	$\pm$	0.60 	&	89.72 	\\
PKS 1124-186                	&	1.048 	&	0.64 	$\pm$	0.34 	&	2.00 	$\pm$	0.04 	&	2.34 	$\pm$	0.05 	&	0.34 	$\pm$	0.05 	&	1.00 	$\pm$	0.24 	&	2.04 	$\pm$	0.49 	&	-5.15 	\\
TXS 0025+197                	&	1.552 	&	0.33 	$\pm$	0.14 	&	2.01 	$\pm$	0.03 	&	2.45 	$\pm$	0.08 	&	0.44 	$\pm$	0.05 	&	1.37 	$\pm$	0.27 	&	3.50 	$\pm$	0.68 	&	-9.37 	\\
PKS 1454-354                	&	1.424 	&	1.57 	$\pm$	0.47 	&	1.92 	$\pm$	0.06 	&	2.45 	$\pm$	0.04 	&	0.53 	$\pm$	0.05 	&	0.66 	$\pm$	0.09 	&	1.59 	$\pm$	0.22 	&	-8.81 	\\
PKS 0202-17                 	&	1.740 	&	0.03 	$\pm$	0.01 	&	2.45 	$\pm$	0.02 	&	3.26 	$\pm$	0.23 	&	0.81 	$\pm$	0.07 	&	2.67 	$\pm$	0.21 	&	7.32 	$\pm$	0.57 	&	10.81 	\\
S4 1030+61                  	&	1.401 	&	0.43 	$\pm$	0.05 	&	2.01 	$\pm$	0.03 	&	2.30 	$\pm$	0.04 	&	0.29 	$\pm$	0.03 	&	1.17 	$\pm$	0.08 	&	2.80 	$\pm$	0.18 	&	-7.59 	\\
S4 0917+44                  	&	2.186 	&	0.26 	$\pm$	0.07 	&	2.13 	$\pm$	0.03 	&	2.67 	$\pm$	0.07 	&	0.54 	$\pm$	0.04 	&	1.41 	$\pm$	0.16 	&	4.50 	$\pm$	0.52 	&	48.16 	\\
OP 313                      	&	0.997 	&	0.16 	$\pm$	0.06 	&	2.00 	$\pm$	0.02 	&	2.53 	$\pm$	0.11 	&	0.53 	$\pm$	0.05 	&	2.17 	$\pm$	0.38 	&	4.33 	$\pm$	0.76 	&	-0.23 	\\
PKS 1004-217                	&	0.330 	&	0.68 	$\pm$	0.15 	&	2.23 	$\pm$	0.04 	&	2.41 	$\pm$	0.04 	&	0.19 	$\pm$	0.04 	&	0.77 	$\pm$	0.07 	&	1.02 	$\pm$	0.09 	&	37.56 	\\
OK 630                      	&	1.446 	&	0.23 	$\pm$	0.09 	&	2.11 	$\pm$	0.02 	&	2.52 	$\pm$	0.07 	&	0.41 	$\pm$	0.04 	&	1.38 	$\pm$	0.23 	&	3.38 	$\pm$	0.57 	&	5.66 	\\
PKS 0226-559                	&	2.464 	&	0.07 	$\pm$	0.03 	&	2.07 	$\pm$	0.02 	&	2.74 	$\pm$	0.18 	&	0.68 	$\pm$	0.06 	&	2.54 	$\pm$	0.49 	&	8.80 	$\pm$	1.69 	&	76.25 	\\
PKS 0035-252                	&	1.196 	&	1.50 	$\pm$	0.27 	&	2.03 	$\pm$	0.04 	&	2.28 	$\pm$	0.03 	&	0.25 	$\pm$	0.03 	&	0.58 	$\pm$	0.05 	&	1.27 	$\pm$	0.11 	&	2.45 	\\
PKS 0420-01                 	&	0.916 	&	0.44 	$\pm$	0.14 	&	2.05 	$\pm$	0.04 	&	2.59 	$\pm$	0.07 	&	0.55 	$\pm$	0.05 	&	1.08 	$\pm$	0.15 	&	2.07 	$\pm$	0.28 	&	10.80 	\\
PKS 0524-485                	&	1.300 	&	0.19 	$\pm$	0.27 	&	2.09 	$\pm$	0.04 	&	2.34 	$\pm$	0.12 	&	0.25 	$\pm$	0.07 	&	1.54 	$\pm$	1.02 	&	3.53 	$\pm$	2.34 	&	-29.60 	\\
PKS 0215+015                	&	1.715 	&	0.01 	$\pm$	0.00 	&	2.09 	$\pm$	0.02 	&	3.01 	$\pm$	0.41 	&	0.93 	$\pm$	0.09 	&	5.19 	$\pm$	0.49 	&	14.10 	$\pm$	1.32 	&	1.20 	\\
TXS 2241+406                	&	1.171 	&	0.06 	$\pm$	0.04 	&	1.99 	$\pm$	0.02 	&	2.31 	$\pm$	0.12 	&	0.32 	$\pm$	0.05 	&	2.87 	$\pm$	0.81 	&	6.23 	$\pm$	1.75 	&	4.34 	\\
B2 2113+29                  	&	1.514 	&	7.39 	$\pm$	1.14 	&	0.88 	$\pm$	0.24 	&	2.39 	$\pm$	0.03 	&	1.51 	$\pm$	0.08 	&	0.31 	$\pm$	0.02 	&	0.77 	$\pm$	0.06 	&	-1.56 	\\
PKS 2320-035                	&	1.393 	&	0.25 	$\pm$	0.16 	&	2.07 	$\pm$	0.04 	&	2.51 	$\pm$	0.11 	&	0.43 	$\pm$	0.06 	&	1.38 	$\pm$	0.40 	&	3.30 	$\pm$	0.95 	&	15.42 	\\
PKS 0116-219                	&	1.165 	&	0.02 	$\pm$	0.00 	&	2.23 	$\pm$	0.02 	&	2.70 	$\pm$	0.21 	&	0.47 	$\pm$	0.07 	&	3.45 	$\pm$	0.34 	&	7.47 	$\pm$	0.74 	&	7.54 	\\
B2 0716+33                  	&	0.779 	&	0.10 	$\pm$	0.08 	&	2.01 	$\pm$	0.03 	&	2.38 	$\pm$	0.16 	&	0.37 	$\pm$	0.07 	&	2.19 	$\pm$	0.91 	&	3.90 	$\pm$	1.61 	&	18.83 	\\
PKS 0405-385                	&	1.285 	&	0.37 	$\pm$	0.19 	&	2.09 	$\pm$	0.04 	&	2.53 	$\pm$	0.08 	&	0.44 	$\pm$	0.06 	&	1.05 	$\pm$	0.24 	&	2.39 	$\pm$	0.54 	&	-6.49 	\\
TXS 1318+225                	&	0.943 	&	0.04 	$\pm$	0.01 	&	2.28 	$\pm$	0.02 	&	2.61 	$\pm$	0.13 	&	0.33 	$\pm$	0.06 	&	2.54 	$\pm$	0.25 	&	4.94 	$\pm$	0.48 	&	-1.01 	\\
OC 457                      	&	0.859 	&	0.51 	$\pm$	0.12 	&	1.99 	$\pm$	0.04 	&	2.39 	$\pm$	0.05 	&	0.40 	$\pm$	0.04 	&	0.94 	$\pm$	0.10 	&	1.74 	$\pm$	0.18 	&	51.10 	\\
4C +14.23                   	&	1.038 	&	0.32 	$\pm$	0.14 	&	1.97 	$\pm$	0.04 	&	2.47 	$\pm$	0.08 	&	0.50 	$\pm$	0.06 	&	1.18 	$\pm$	0.24 	&	2.41 	$\pm$	0.48 	&	17.90 	\\
PKS 0131-522                	&	0.925 	&	0.64 	$\pm$	0.13 	&	2.00 	$\pm$	0.04 	&	2.38 	$\pm$	0.04 	&	0.38 	$\pm$	0.04 	&	0.77 	$\pm$	0.07 	&	1.49 	$\pm$	0.14 	&	3.03 	\\
PKS 1441+25                 	&	0.939 	&	0.21 	$\pm$	0.06 	&	1.87 	$\pm$	0.03 	&	2.26 	$\pm$	0.06 	&	0.40 	$\pm$	0.04 	&	1.51 	$\pm$	0.22 	&	2.93 	$\pm$	0.43 	&	-12.15 	\\
PKS 0514-459                	&	0.194 	&	0.30 	$\pm$	0.14 	&	2.03 	$\pm$	0.05 	&	2.42 	$\pm$	0.07 	&	0.39 	$\pm$	0.06 	&	1.09 	$\pm$	0.24 	&	1.31 	$\pm$	0.28 	&	-0.48 	\\
B2 1732+38A                 	&	0.976 	&	0.48 	$\pm$	0.38 	&	1.95 	$\pm$	0.10 	&	2.51 	$\pm$	0.10 	&	0.56 	$\pm$	0.10 	&	0.92 	$\pm$	0.32 	&	1.82 	$\pm$	0.64 	&	9.57 	\\
PKS 0244-470                	&	1.385 	&	0.12 	$\pm$	0.07 	&	2.24 	$\pm$	0.04 	&	2.81 	$\pm$	0.14 	&	0.57 	$\pm$	0.07 	&	1.45 	$\pm$	0.34 	&	3.45 	$\pm$	0.82 	&	17.99 	\\
S4 1849+67                  	&	0.657 	&	0.76 	$\pm$	0.21 	&	1.71 	$\pm$	0.06 	&	2.40 	$\pm$	0.05 	&	0.69 	$\pm$	0.06 	&	0.76 	$\pm$	0.10 	&	1.26 	$\pm$	0.17 	&	65.45 	\\
S5 1217+71                  	&	0.451 	&	0.58 	$\pm$	0.13 	&	1.73 	$\pm$	0.08 	&	2.51 	$\pm$	0.06 	&	0.87 	$\pm$	0.06 	&	0.75 	$\pm$	0.08 	&	1.08 	$\pm$	0.12 	&	15.03 	\\

\hline
\multicolumn{9}{c}{Average}\\
\hline
Mean. / std.&1.14 / 0.56&1.35 / 2.39 & 2.12 / 0.23 & 2.56 / 0.25 & 0.45 / 0.19 & 1.33 / 0.71 & 2.90 / 1.92 &\\
\hline
\end{tabular}}
\end{adjustwidth}
\end{center}
{\footnotesize{{Note. 
 \\
Column (3) is in unit of $\rm 10^{-11}\ cm^{-2}s^{-1}MeV^{-1}$; \\
Column (7) and (8) are in unit of $\rm GeV$.\\

}}}
\end{table}

\section{Comparison of fitting of two models} \label{sec:comparison}

Figure \ref{fig:fig1} shows the fitting results of the default models (LP) and the BPL model for the gamma-ray spectra of 87 bright FSRQs. To highlight the subtle differences between the fit lines of the two models, we calculate the fractional difference between the models' best-fitted lines:
\begin{equation}
{\rm Frac\ Diff}\ (E)=\frac{S_{\rm LP} (E) - S_{\rm BPL}(E)}{S_{\rm BPL}(E)} \text {. }
\end{equation}
Here, $S_{\rm LP}(E)$ represents the predicted SED at energy E for the LP model, and $S_{\rm BPL}(E)$ represents the predicted SED at energy E for the BPL model. The fractional difference between the two models is indicated by the blue area in Figure \ref{fig:fig1}. From Figure \ref{fig:fig1}, it is apparent that there is overlap between the two models in most energy ranges, with differences mainly concentrated near the upper and lower energy limits of the Fermi spectral analysis or around the spectral break energy.

We perform an Akaike Information Criterion (AIC) test to evaluate the goodness of fit, determining the AIC value for each model as described by \citep{1974ITAC...19..716A}:
\begin{equation}\label{eq5}
{\rm AIC} = 2k - {2\ln{\mathcal L}} \text {. }
\end{equation}
Where, $\rm {{\mathcal L}}$ is the maximum value of the likelihood function for the model being evaluated, and $k$ denotes the number of free parameters in the model. A model with a smaller AIC is considered to have a better fit. The relative goodness of fit between model s and model $\rm s^{\prime}$ can be evaluated by the difference in AIC values:
\begin{equation}\label{eq6}
\Delta \mathrm{AIC}_{s, s^{\prime}}=\mathrm{AIC}s-\mathrm{AIC}{s^{\prime}} \text {. }
\end{equation}
The $\rm \Delta \mathrm{AIC}_{s, s^{\prime}}$ estimates how much more model s diverges from the true distribution compared to model $\rm s^{\prime}$, also known as the relative Kullback-Leibler information quantities of the two models (e.g., \citealt{RN52,2012ApJ...761....2H, 2021MNRAS.500.5297A}). When $\rm \Delta AIC > 10$, it indicates that the difference between models is highly significant; The model with the lower AIC value has strong support, while the model with the higher AIC value can essentially be excluded.

The BPL model has one more free parameter than the LP model. The difference in AIC between the two models is given by $\rm \Delta \mathrm{AIC}=\mathrm{AIC}{\scriptscriptstyle BPL}-\mathrm{AIC}{\scriptscriptstyle LP} = 2- 2(ln{\mathcal L}{\scriptscriptstyle BPL}-ln{\mathcal L}{\scriptscriptstyle LP})$. Here, $\rm {{\mathcal L}{\scriptscriptstyle BPL}}$ is the maximum value of the likelihood function for the BPL model, and $\rm {{\mathcal L}{\scriptscriptstyle LP}}$ is the maximum value of the likelihood function for the LP model. The $\rm \Delta \mathrm{AIC}$ values for the 87 FSRQs are tabulated in Table \ref{tab:3}. We find that the 87 bright FSRQs can be divided into three groups:
\begin{itemize}	
 \item [1.]
LP-preferred FSRQs: The $\rm\Delta \mathrm{AIC}$ of 38 samples is larger than 10, indicating a clear preference for the LP model.
 \item [2.]
BPL-preferred FSRQs: The $\rm\Delta \mathrm{AIC}$ of 20 FSRQs is smaller than -10, indicating strong evidence supporting the superiority of the BPL model.
 \item [3.]
FSRQs with comparable model fitting: The $\rm\Delta \mathrm{AIC}$ of the remaining 29 FSRQs is between -10 and 10, indicating that their gamma-ray spectra cannot exclude either LP model or the BPL model.
\end{itemize}	

\renewcommand\thefigure{2}
\begin{figure*}[!htp]
\centering
\includegraphics[height=6cm,width=8cm]{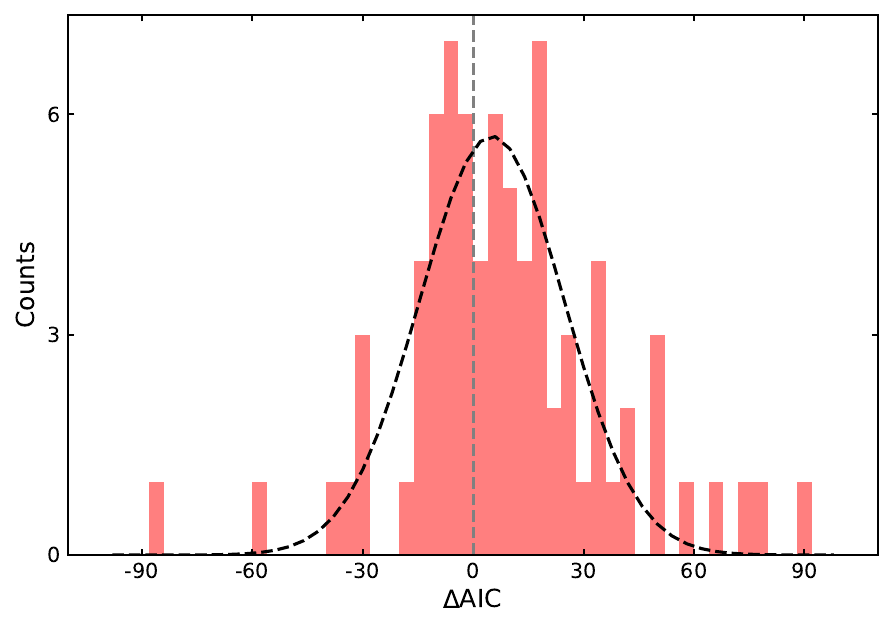}
\caption{The distribution of $\rm\Delta \mathrm{AIC} $ in 87 bright FSRQs. The black dashed line represents the Gaussian fitting line.}
\label{fig:fig2}
\end{figure*}

The gamma-ray fitting of FSRQs shows complex preferences for both BPL and LP spectra, consistent with previous researches (e.g., \citealt{2012ApJ...761....2H,2014MNRAS.441.3591H}). The distribution of $\rm\Delta \mathrm{AIC}$ for the 87 bright FSRQs is presented in Figure \ref{fig:fig2}. Fitting this distribution with a Gaussian function, which yields an expected value close to 0, indicates that the population of FSRQs exhibits comparable preferences for these two types of models.

To explore potential intrinsic physical differences between LP-preferred FSRQs and BPL-preferred FSRQs, we study the distribution of these two groups in a multidimensional parameter space and conducted a clustering analysis (for more details, see Appendix \ref{sect:appendix a}). Our analysis revealed no significant differences in the parameter distributions between the two types of sources. The results from the clustering analysis suggest that LP-preferred FSRQs and BPL-preferred FSRQs likely belong to the same category, indicating that the gamma-ray spectra of these samples probably share a common physical origin.

Notably, for some sources, preferences between the BPL and LP models can shift across different time periods \citep{2012ApJ...761....2H,2014MNRAS.441.3591H}. This implies the absence of rigid boundaries distinguishing the BPL and LP models, suggesting their potential equivalence.

These findings support the argument for the equivalence of the BPL and LP models in modeling FSRQ gamma-ray spectra within the 0.1-10 GeV energy range. We suggest that the LP and BPL models can be considered approximations of each other, with the evolution of the spectral index in the LP model indicating a smooth GeV spectral break. The choice between the BPL and LP models may reflect the nature of the GeV spectral break, whether sharp or smooth. In such cases, using the BPL model to locate the GeV spectral break energy is effective for all FSRQs.

\section{Results and Discussion} \label{sec:results and disscussion}
\subsection{GeV Spectral break}

All 87 samples show significant softening ($\Delta \gamma > 0.1$) in the gamma-ray spectra fitting, indicating that spectral breaks are a common phenomenon in FSRQs. In the observed frame, the spectral break energy ranges from 0.31 to 5.19 GeV, with an average of $\langle E_b \rangle = 1.33 \pm 0.71$ GeV. This shifts to $\langle E_b^{\prime} \rangle = 2.90 \pm 1.92$ GeV in the rest frame. Additionally, the low-energy photon index of the 87 bright FSRQs ranges from 0.88 to 2.70, with an average value of $\langle \gamma_1 \rangle = 2.12 \pm 0.23$; the high-energy photon index ranges from 2.07 to 3.62, with an average value of $\langle \gamma_2 \rangle = 2.56 \pm 0.25$; and the change in photon index ranges from 0.19 to 1.51, with an average value of $\langle \Delta \gamma \rangle = 0.45 \pm 0.19$.

The gamma-ray flux differences among FSRQs are substantial. To study their gamma-ray spectral shape and evolution, we normalize the BPL-fitted gamma-ray spectra of the 87 FSRQs with the differential energy flux at the spectral break energy, $\nu F_\nu (E_b)$, represented by gray lines in Figure \ref{fig:fig3}. The left panel of Figure \ref{fig:fig3} shows the normalized spectra in the observed frame, while the right panel shows the normalized spectra in the rest frame. Using Monte Carlo sampling, we estimate the average gamma-ray spectra (red lines) and provide the 1-$\sigma$ confidence region (enclosed by the red dotted lines). Figure \ref{fig:fig3} clearly displays the spectral break phenomenon in FSRQs.

\renewcommand\thefigure{3}
\begin{figure*}
\centering
\includegraphics[height=8.5cm,width=8.5cm]{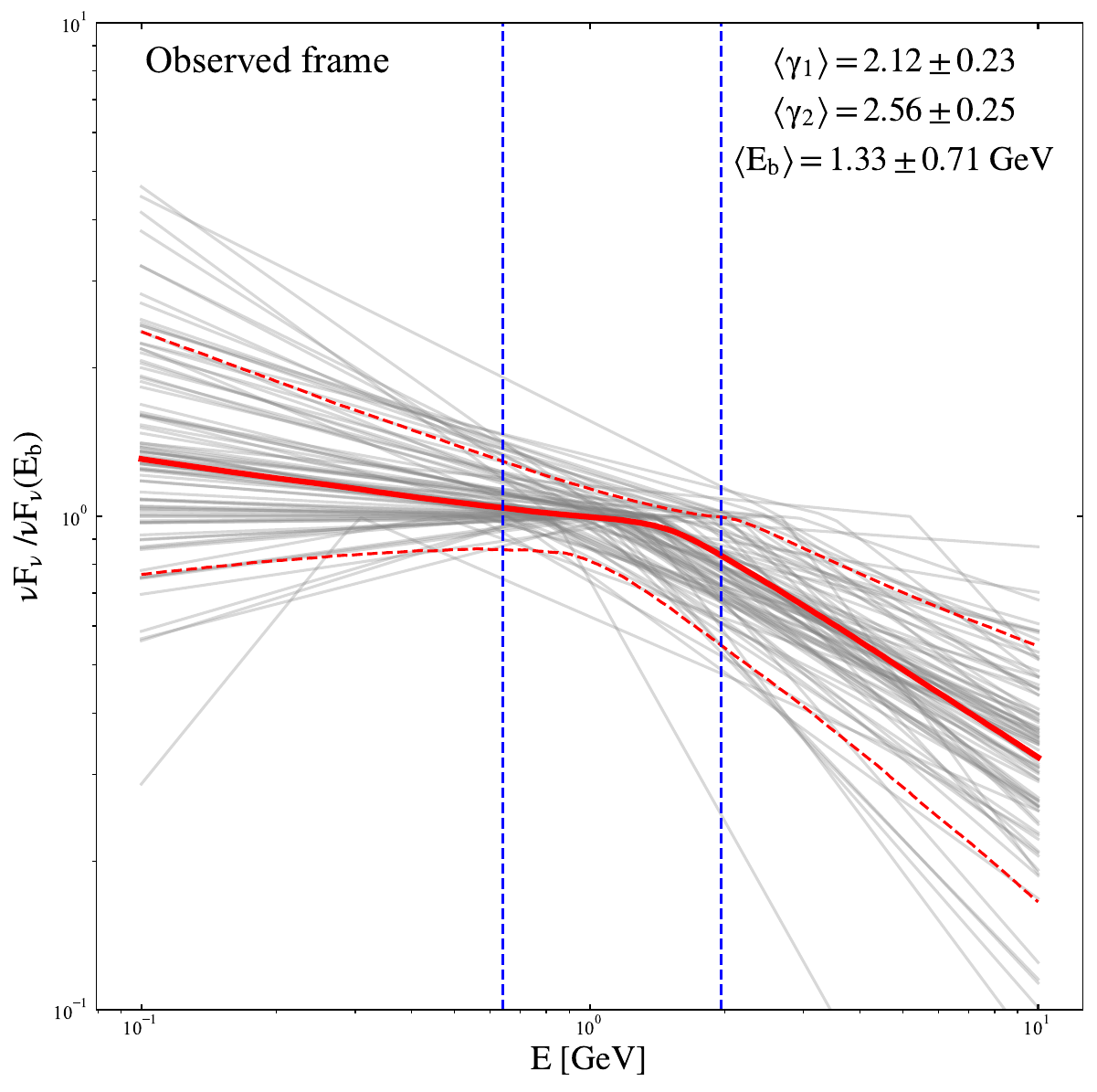}
 \includegraphics[height=8.5cm,width=8.5cm]{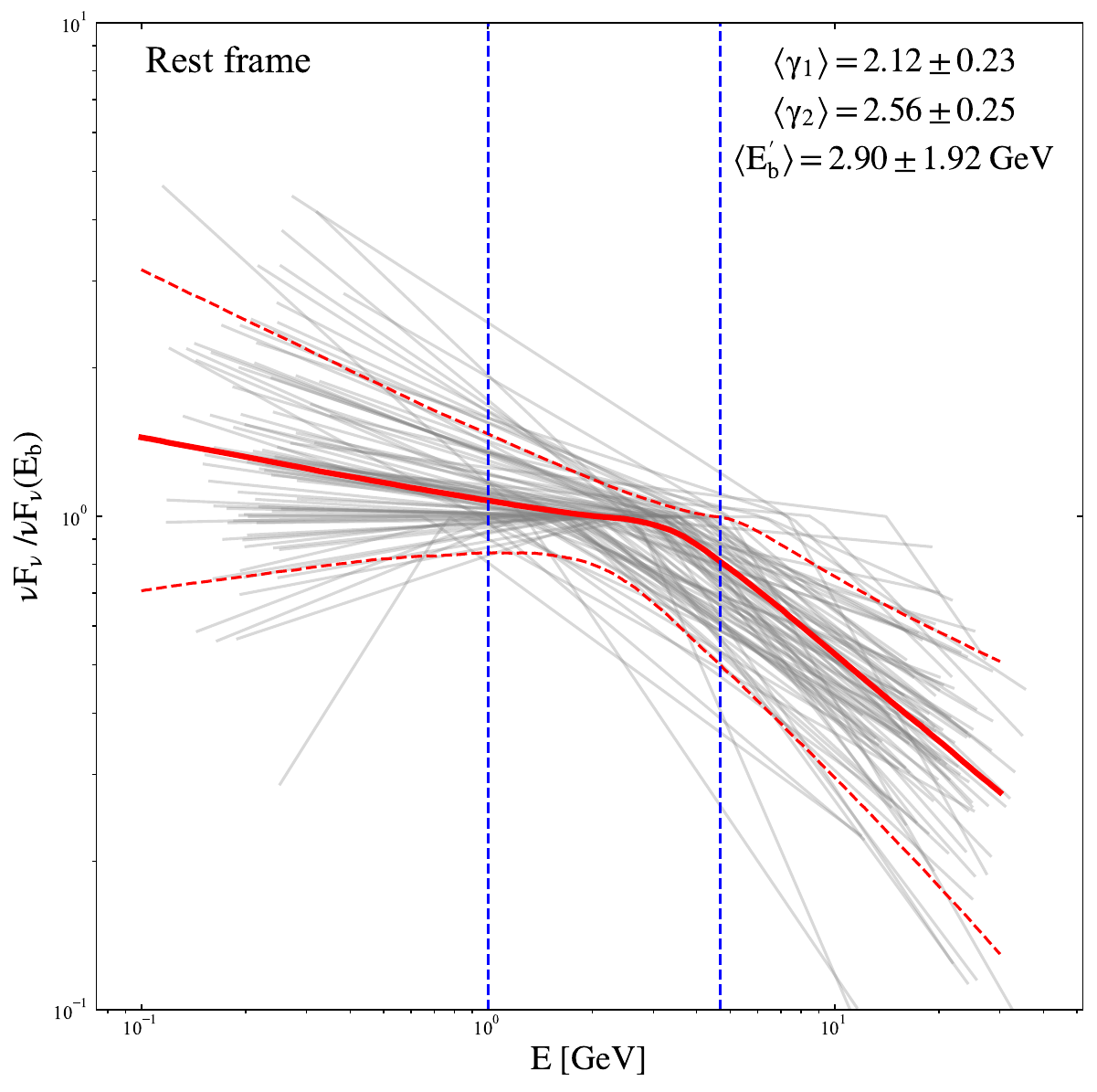}
 
 \caption{The normalized energy spectra of 87 bright FSRQs in the observed frame (left panel) and their rest frames (right panel). the grey lines represent the normalized gamma-ray spectra of individual samples, the red lines represent the average gamma-ray spectra estimated through Monte Carlo sampling, and the red dashed lines indicate the 1-$\sigma$ confidence region. The blue dashed lines represent the spectral break energy range. 
}
 \label{fig:fig3}
\end{figure*}

The distribution of some BPL fitting parameters for the 87 FSRQs is presented in Figure \ref{fig:fig4}. Panel (a) displays the relationship between the spectral break energy and redshift; panel (b) illustrates the relationship between the low-energy spectral index and the high-energy spectral index; and panel (c) shows the relationship between the change in spectral index and redshift. We find a strong correlation between the low-energy and high-energy spectral indices, with a Spearman correlation coefficient of 0.79 and a p-value of $\rm 1.32 \times 10^{-19}$. No correlation was found between redshift and either break energy or the change in spectral index.

\subsection{Comparison with other fitting results}

Several studies have employed BPL models to fit the gamma-ray spectra of FSRQs. For example, \cite{2010ApJ...710.1271A} conducted BPL fitting on 12 bright FSRQs spanning the energy range from 100 MeV to several tens of GeV. They observed break energies in the rest frame ranging from 1.61 to 9.91 GeV, with an average value of $\langle E_b^{\prime} \rangle = 4.89 \pm 2.71$ GeV. \cite{2010ApJ...717L.118P} utilized BPL models and PL plus double absorption models for recombination continua of hydrogen and $\rm He\,{\sc ii}$ to fit the gamma-ray spectra of 9 bright FSRQs across the energy range from 100 MeV to 100 GeV. They observed break energies in the rest frame ranging from 1.80 to 19 GeV, with an average value of $\langle E_b^{\prime} \rangle = 6.41 \pm 5.07$ GeV. Furthermore, \cite{2012ApJ...761....2H} used BPL models to search for evidence of breaks in the spectrum at both low energies (ranging from a few hundred MeV to 12 GeV) and high energies (after the first break energy) to validate the double absorption model. In the low-energy fitting, they observed break energies in the rest frame ranging from 1.3 to 6 GeV, with an average value of $\langle E_b^{\prime} \rangle = 3.13 \pm 1.51$ GeV.

We compare our fitting results with those of previous studies in Figure \ref{fig:fig4}. In Figure \ref{fig:fig4}, the black points represent the BPL model parameters of 87 FSRQs in the present work; the red points represent 12 sources from \cite{2010ApJ...710.1271A}; the blue points represent 9 bright FSRQs from \cite{2010ApJ...717L.118P}; and the green points represent the low-energy fitting results of 9 bright FSRQs in \cite{2012ApJ...761....2H}.

\renewcommand\thefigure{4}
\begin{figure*}
\centering
\includegraphics[height=8.5cm,width=8.5cm]{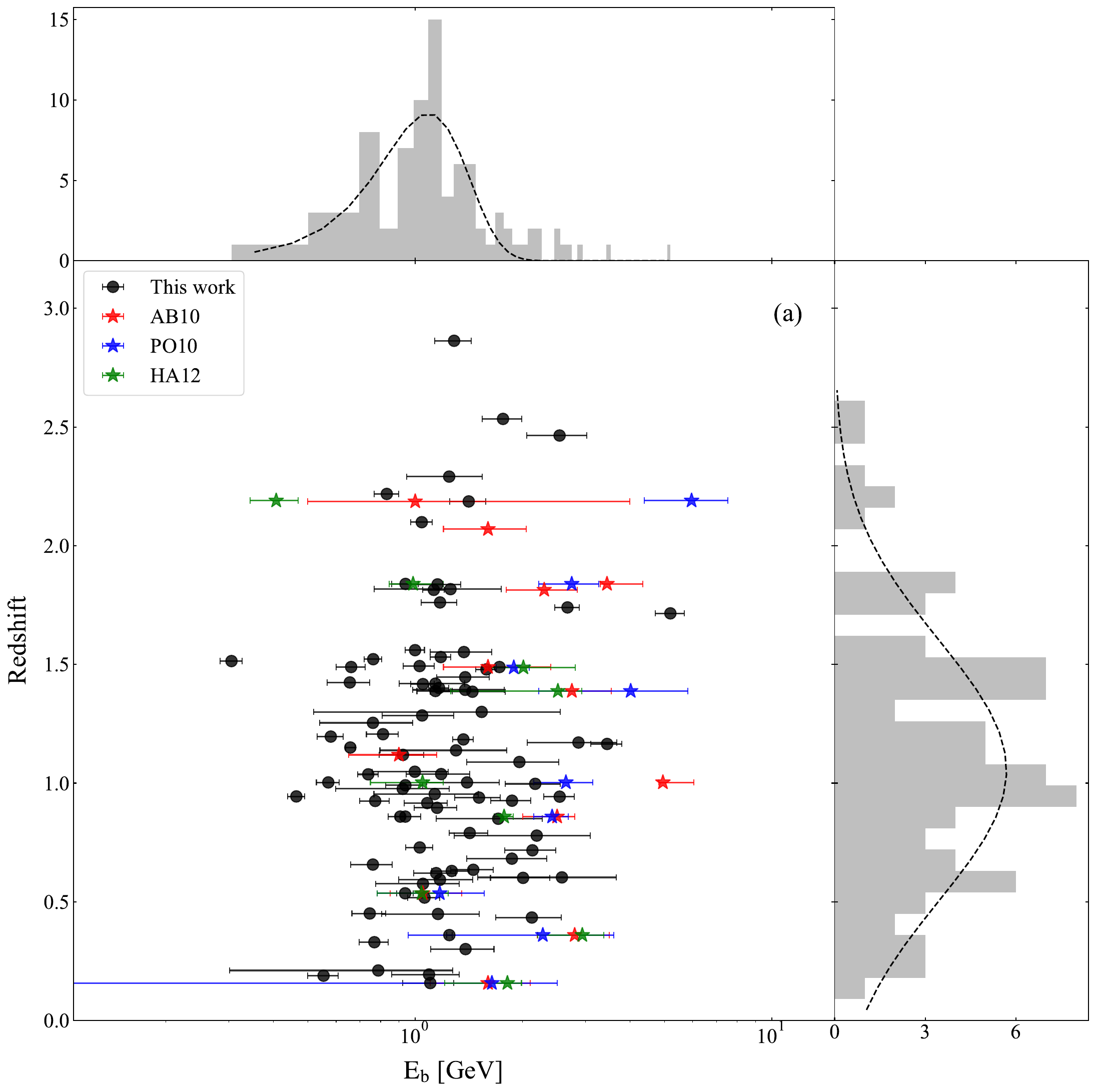}
 \includegraphics[height=8.5cm,width=8.5cm]{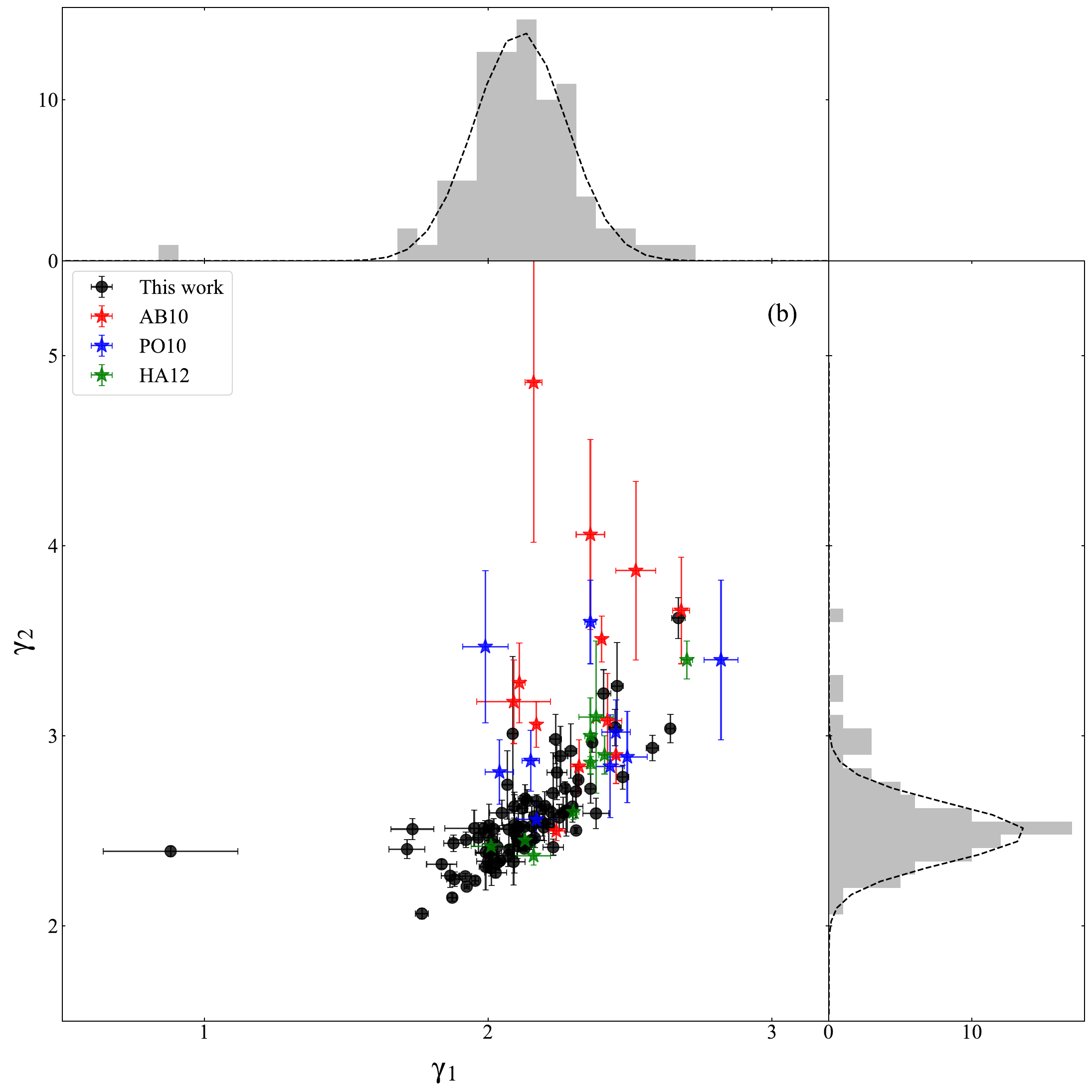}
  \includegraphics[height=8.5cm,width=8.5cm]{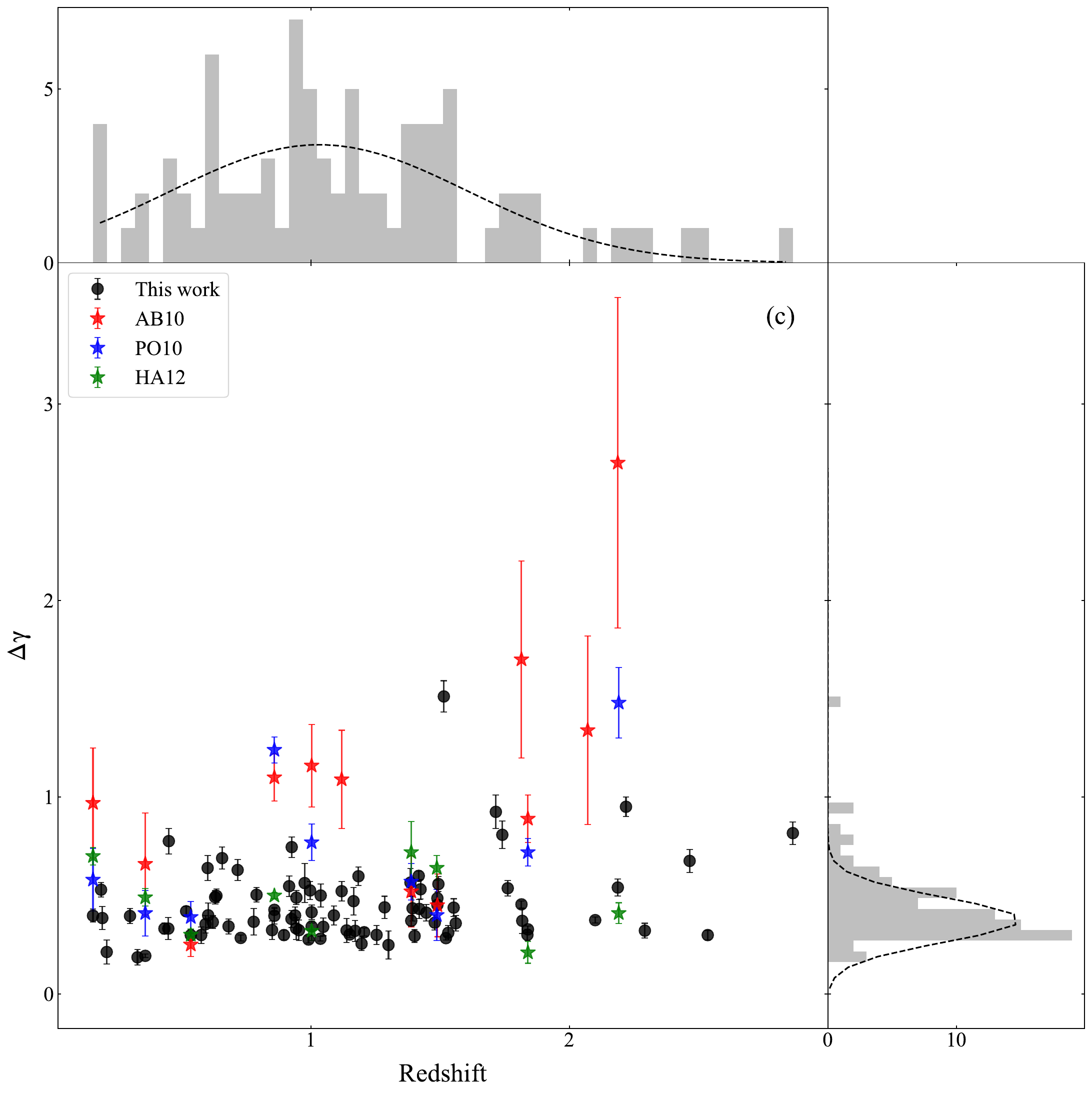} 
  \caption{Distribution of part BPL fitting parameter: (a) the spectral break energy and the redshift; (b) the low energy spectral index and the high energy spectral index; (c) the redshift and the change in spectral index. The black points represent 87 bright FSRQs in this study; the red points show \cite{2010ApJ...710.1271A}'s fitting results for 12 bright FSRQs; the blue points represent \cite{2010ApJ...717L.118P}'s fitting results for 9 bright FSRQs; green points are \cite{2012ApJ...761....2H}'s fitting results for 9 bright FSRQs.}
 \label{fig:fig4}
\end{figure*}

From the figure, we find that:
i) Our fitting, compared to previous work, employs data with longer exposure times, resulting in smaller uncertainties in the fitted parameters.
ii) Our fitting results are more consistent with the distribution of low-energy fitting results of \cite{2012ApJ...761....2H}.
iii) Compared with the fitting results from \cite{2010ApJ...710.1271A} and \cite{2010ApJ...717L.118P}, we obtain smaller break energies ($E_b$), smaller high-energy spectral indices ($\gamma_2$), and smaller changes in photon index ($\Delta \gamma$).
iv) The correlation between redshift and the change in spectral index found in the fitting results of \cite{2010ApJ...710.1271A} and \cite{2010ApJ...717L.118P} is likely due to the inclusion of data above 10 GeV, introducing the effect of EBL absorption on the gamma-ray spectra. However, this correlation is not observed in our study and the work of \cite{2012ApJ...761....2H}. 
This difference may arise from the different energy ranges chosen for the fitting process.

\renewcommand\thefigure{5}
\begin{figure*}
\centering
\includegraphics[height=6cm,width=8cm]{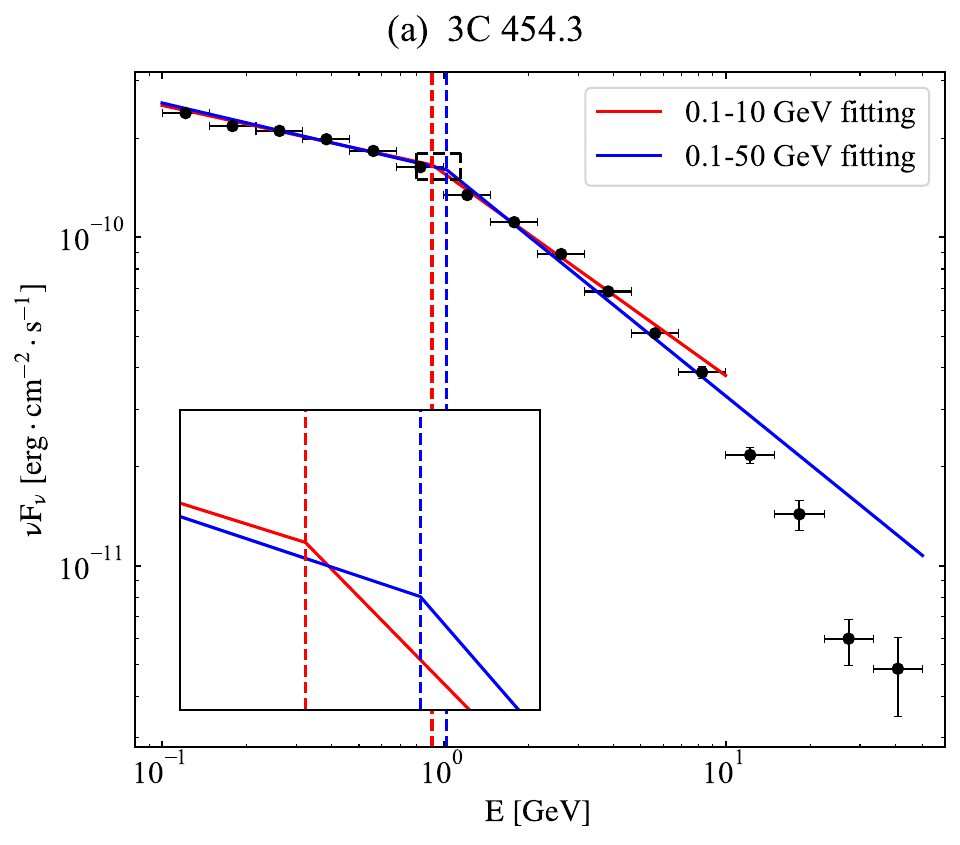}
 \includegraphics[height=6cm,width=8cm]{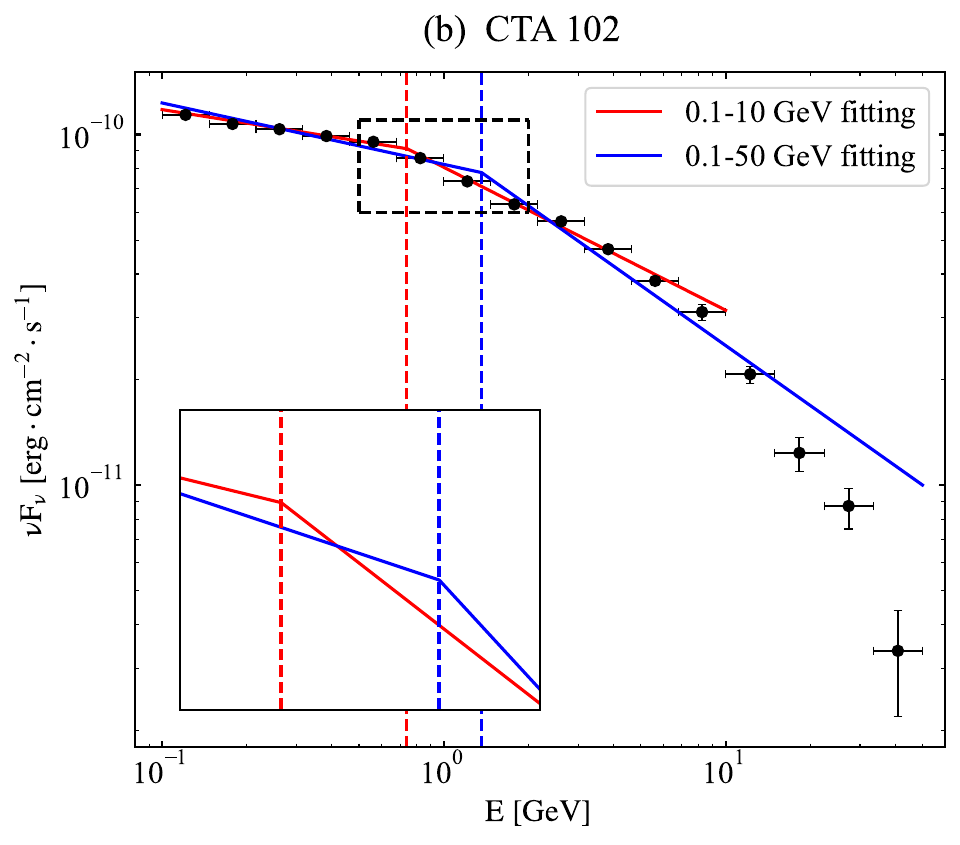}
  \includegraphics[height=6cm,width=8cm]{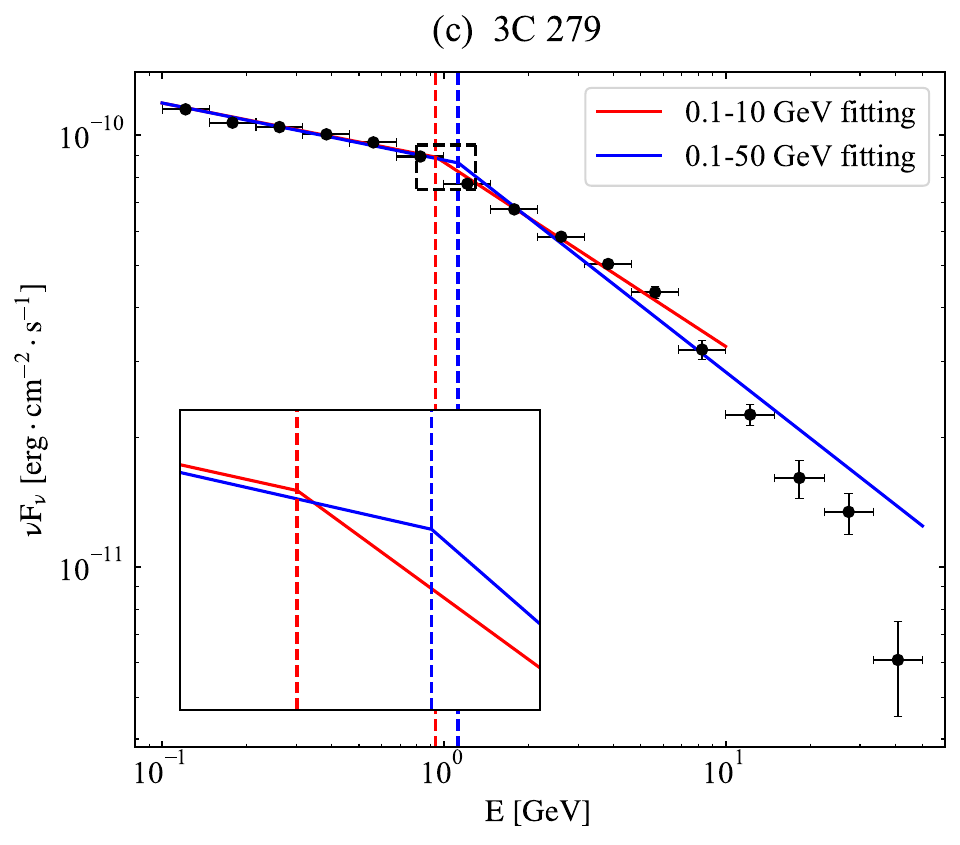}
   \includegraphics[height=6cm,width=8cm]{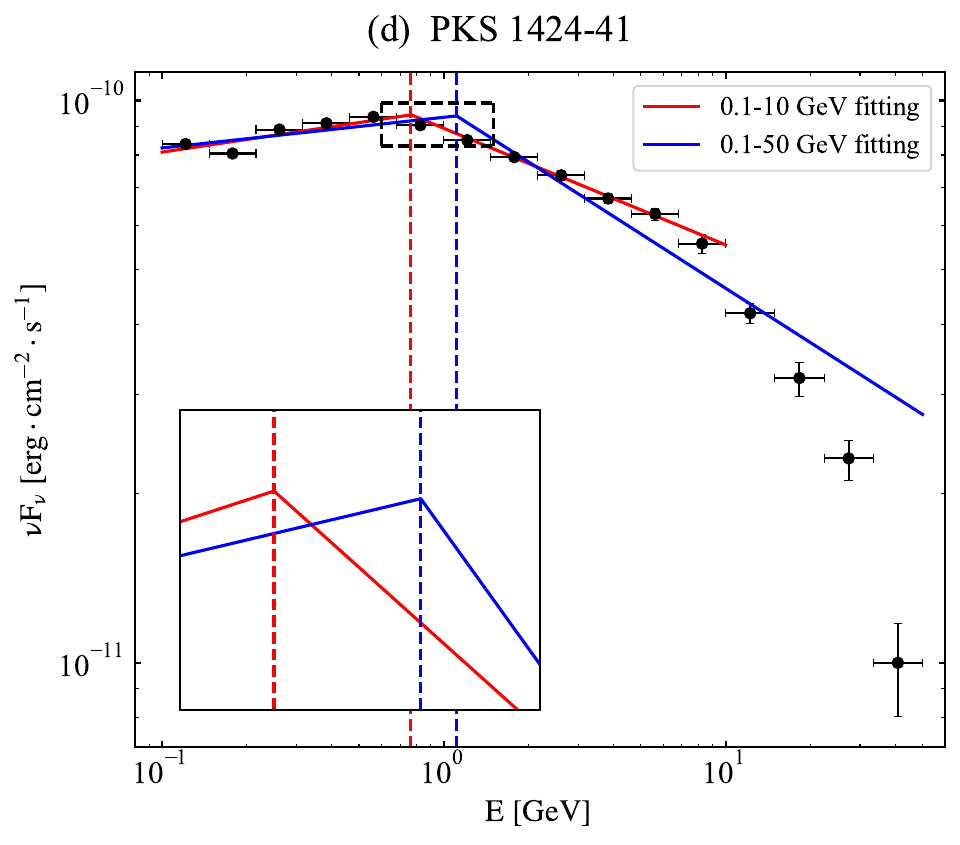}
  \caption{The BPL fitting results for the four brightest FSRQs in the 0.1-10 GeV range (red line) and the 0.1-50 GeV range (blue line) are depicted for: (a) 3C 454.3; (b) CTA 102; (c) 3C 279; (d) PKS 1421-41. The inset provides a detailed view near the spectral break energy.}
 \label{fig:fig5}
\end{figure*}

We extend the spectral fitting energy upper limit to 50 GeV for four highly significant FSRQs, namely 3C 454.3, CTA 102, 3C 279, and PKS 1424-41, and performed BPL fitting. Figure \ref{fig:fig5} illustrates the fitting results of the BPL model for these four FSRQs across different energy ranges. The black dots represent the actual SED data; the red line depicts the best fit of the BPL model in the 0.1-10 GeV range, while the blue line represents the best fit in the 0.1-50 GeV range. The fitting parameters are listed in Table \ref{tab:5}.

In the Figure \ref{fig:fig5}, it is evident that the BPL model provides a good fit to the gamma-ray spectra below 10 GeV (smaller $\rm \Delta AIC$). However, around 10 GeV, the spectra of all four FSRQs exhibit a faster cutoff than a power law. The BPL model fails to cover this spectral range. Consequently, in the overall fitting in the 0.1-50 GeV range, the BPL model tends to have a higher break energy and a softer high-energy power law component to accommodate the data. This trend is consistent with the fitting results presented in \cite{2010ApJ...710.1271A} and \cite{2010ApJ...717L.118P}. However, this results in deviations from the actual data, leading to a rapid decrease in the goodness of fit of the BPL model (see $\rm \Delta AIC$ in Table \ref{tab:5}).

The spectral cutoff at 10 GeV may arise from the absorption of photons in the broad-line region (e.g., \citealt{2006ApJ...653.1089L, 2008ApJ...688..148L}) or from the absorption of extragalactic background photons (e.g., \citealt{2005ApJ...618..657D, 2017A&A...603A..34F}), but it is not yet fully understood. Including the spectra above 10 GeV in the fitting range will pose challenges for accurately locating the energy of the GeV spectral break.

Above 10 GeV, the simple BPL model is not applicable, and an exponential cutoff or a super-exponential cutoff is needed. As the exposure time of Fermi-LAT increases, the resolution of the gamma-ray spectrum improves, allowing more complex mathematical models to be used for accurately fitting the gamma-ray spectrum of blazars. However, the potential physical mechanisms are not entirely clear. The focus of our future research will be to determine whether the LP model of FSRQs' gamma-ray spectra is more of a mathematical approximation of the BPL with an exponential cutoff or whether it better reflects the intrinsic curvature characteristics of the spectrum.

\renewcommand\thetable{4}
\begin{table}
\begin{center}
\caption{Comparison of BPL fitting results for different fitting energy band}
\label{tab:4}
\begin{adjustwidth}{-2.2cm}{-5cm}
\resizebox{20cm}{!}{
\begin{tabular}{ccccccccccccc} \hline
{Source Name}& Label & $E_b$&\multirow{2}{*}{Trends} & $\gamma_{\rm 1}$&\multirow{2}{*}{Trends}& $\gamma_{\rm 2}$&\multirow{2}{*}{Trends}&$\Delta \gamma$&\multirow{2}{*}{Trends} &$\rm \Delta AIC$&\multirow{2}{*}{Trends}  \\
\normalsize(1) & \normalsize(2) & \normalsize(3) &&\normalsize(4) && \normalsize(5) &&  \normalsize(6) & & \normalsize(7)   \\
\hline 
\multirow{2}{*}{3C 454.3}& 0.1-10 GeV fitting & $0.91\pm0.03$ &\multirow{2}{*}{$\uparrow$}& $2.19\pm0.01$ &\multirow{2}{*}{$\uparrow$}&$2.62\pm0.01$ &\multirow{2}{*}{$\uparrow$}&$0.42\pm0.01$ &\multirow{2}{*}{$\uparrow$}&-1305.32&\multirow{2}{*}{$\uparrow$}\\
& 0.1-50 GeV fitting&$1.02\pm0.02$&&$2.20\pm0.01$&&$2.69\pm0.01$&&$0.49\pm0.01$&&345.16&\\
\multirow{2}{*}{CTA 102}& 0.1-10 GeV fitting& $0.74\pm0.05$ &\multirow{2}{*}{$\uparrow$}&$2.13\pm0.01$ &\multirow{2}{*}{$\uparrow$}&$2.41\pm0.01$ &\multirow{2}{*}{$\uparrow$}&$0.28\pm0.01$ &\multirow{2}{*}{$\uparrow$}&-2033.15&\multirow{2}{*}{$\uparrow$}\\
& 0.1-50 GeV fitting&$1.37\pm0.06$&&$2.17\pm0.01$&&$2.57\pm0.01$&&$0.40\pm0.01$&&122.37&\\
\multirow{2}{*}{3C 279}& 0.1-10 GeV fitting& $0.94\pm0.05$ &\multirow{2}{*}{$\uparrow$}&$2.13\pm0.01$ &\multirow{2}{*}{$\uparrow$}&$2.43\pm0.01$ &\multirow{2}{*}{$\uparrow$}&$0.30\pm0.01$ &\multirow{2}{*}{$\uparrow$}&-1529.33&\multirow{2}{*}{$\uparrow$}\\
& 0.1-50 GeV fitting&$1.37\pm0.06$&&$2.17\pm0.01$&&$2.57\pm0.01$&&$0.40\pm0.01$&&84.27&\\
\multirow{2}{*}{PKS 1424-41}& 0.1-10 GeV fitting& $0.76\pm0.04$ &\multirow{2}{*}{$\uparrow$}&$1.92\pm0.01$ &\multirow{2}{*}{$\uparrow$}&$2.21\pm0.01$ &\multirow{2}{*}{$\uparrow$}&$0.28\pm0.01$ &\multirow{2}{*}{$\uparrow$}&-873.91&\multirow{2}{*}{$\uparrow$}\\
& 0.1-50 GeV fitting&$1.11\pm0.05$&&$1.94\pm0.01$&&$2.32\pm0.01$&&$0.38\pm0.01$&&222.88&\\

\hline

\end{tabular}}
\end{adjustwidth}
\end{center}
{\footnotesize{Note. Column (1) is source name;
Column (2) labels the Fermi spectral analysis energy band;
Column (4) - (10) are the fitting parameters;
Trends Column show the evolution trends of fitting parameters in different fitting energy band.
 \\}}
\end{table}

\subsection{The origin of GeV spectral break}
\subsubsection{Absorption by emission lines from the BLR}

The origin of the GeV spectral break includes gamma-ray absorption via photon-photon pair production from emission lines in the BLR. \cite{2010ApJ...717L.118P} simulated the BLR environment and estimated the energy at which strong lines cause a break. Photons from the BLR with an energy $E_{line}(eV)$ can absorb gamma-ray photons, with the optical depth given by: 
\begin{equation}\label{eq8}
\rm \tau (E_\gamma , E_{line})= \tau_T \frac{\sigma_{\gamma \gamma}(s)}{\sigma_{T}}.
\end{equation}
Here, $s=E_\gamma E_{line}/{(m_e c^2)}^2$, and ${\sigma_{\gamma \gamma}}/{\sigma_{T}}$ represents the gamma-gamma pair production cross section relative to the Thomson scattering cross section. The line optical depth $\tau_T$ is given by \citep{2010ApJ...717L.118P}:
\begin{equation}\label{eq9}
\tau_\mathrm{T}=\frac{L_{\rm line}\sigma_\mathrm{T}}{4\pi R_{\rm line}cE_{\rm line}}.
\end{equation}
In this equation, $\rm L_{\rm line}$ denotes the line luminosity, and $\rm R_{\rm line}$ is the size of the emission line region. The optical depth experiences a jump at the threshold energy $\rm E_\gamma=261/(E_{\mathrm{line}}\left(\mathrm{eV}\right)[1+z])\ \mathrm{GeV}$, leading to the formation of a photon spectral break above this energy (See \citealt{2010ApJ...717L.118P, 2011MNRAS.417L..11S}). The change in spectral index is approximately $\Delta\gamma\approx\tau_\mathrm{T}/4$. 

According to \cite{2010ApJ...717L.118P}, five strong lines and recombination continua from the BLR are potential origins of the observed GeV break: 
1) High-ionization He\,{\sc ii} complex (40.8 - 54.4 eV); 
2) Hydrogen Ly lines and recombination continuum (8.0 - 13.6 eV); 
3) Low-ionization He\,{\sc i} 20 eV complex (21.2 - 24.6 eV); 
4) Low-ionization O\,{\sc vii} and C\,{\sc v} (305 - 560 eV); 
5) High-ionization O\,{\sc viii} and C\,{\sc vi} (367 - 774 eV).

\renewcommand\thefigure{6}
\begin{figure}
\centering
 \includegraphics[height=8cm,width=10cm]{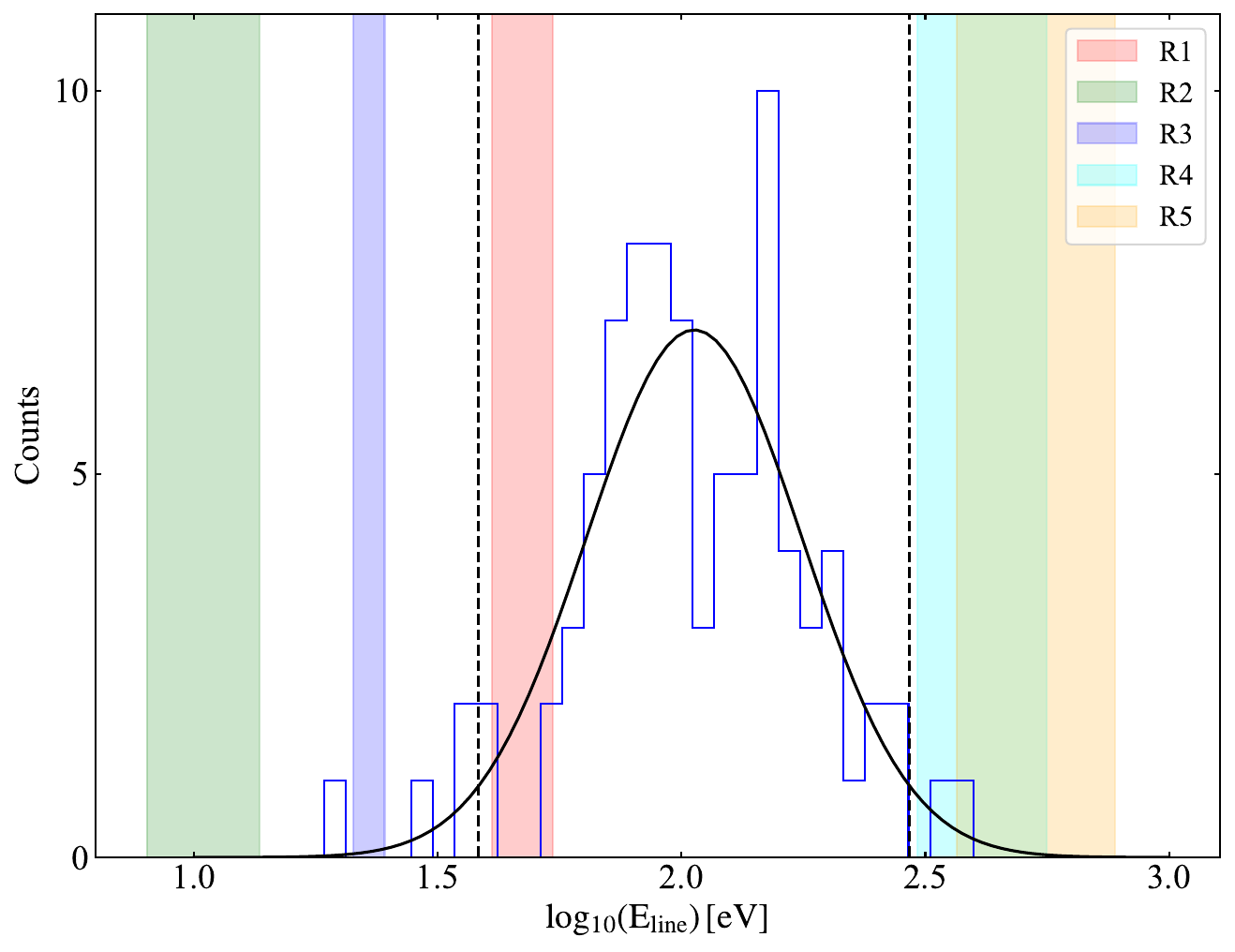}
\caption{Energy distribution of simulated BLR emission line photons. The black solid line represents the Gaussian fit to the energy distribution, and the black dashed lines indicate the 2-sigma confidence interval.  The colored regions highlight the five strongest emission lines and recombination continua identified in \cite{2010ApJ...717L.118P}, specifically: 
R1: High-ionization He\,{\sc ii} complex;
R2: Hydrogen Ly lines and recombination continuum;
R3: Low-ionization He\,{\sc i} 20 eV complex;
R4: Low-ionization O\,{\sc vii} and C\,{\sc v};
R5: High-ionization O\,{\sc viii} and C\,{\sc vi}.   }
 \label{fig:fig6}
\end{figure}

Using the break energy obtained in this context,, we provide the``expected" BLR emission line photon energies, as shown in Figure \ref{fig:fig6}. We fit the energy distribution using a Gaussian function and find that: 
i) the expected emission line photon energy is approximately 100 eV; 
ii) the high-ionization He\,{\sc ii} complex line lies outside the 1-$\sigma$ confidence region of our sample but within the 2-$\sigma$ confidence region; 
iii) the remaining four emission lines are all outside the 2-$\sigma$ confidence region. At a 94.8$\%$ confidence level, these four emission lines can be confidently excluded as the origin of the observed GeV break, while the He\,{\sc ii} complex line remains a plausible candidate.

By fixed the absorption energy, \cite{2010ApJ...717L.118P} and \cite{2011MNRAS.417L..11S} employed a power-law plus double absorption model to fit the GeV break in several bright FSRQs. Their results suggest that absorption from the He\,{\sc ii} complex is a plausible origin for the GeV break, with the optical depth $\rm \tau_{He}$ ranging from 0 to 6.1. \cite{2011ApJ...733...19T} estimated the absorption optical depth of FSRQ \emph{4C +21.35} using the BLR luminosity and found that $\rm \tau_{He}$ is about 2, consistent with the observed spectral index change of $\rm \Delta\gamma \sim 0.5$. 

The energy of He\,{\sc ii} complex photon is about 50 eV. The line luminosity depends on the accretion disk luminosity, given by $\rm L_{He}\simeq0.1 \times L_{Ly\alpha} \simeq  0.018 \times L_{BLR} \simeq 0.0018 \times {L_{disk}}\ erg/s$ (e.g., \citealt{2011ApJ...733...19T, 2021ApJS..253...46P}).  The size of the high-ionization region in the BLR also depends on the disk luminosity, where $\rm R_{He} \simeq 0.5\times R_{BLR}$ \citep{2011ApJ...733...19T} and $\rm R_{BLR} \simeq 0.1 \times ({{L_{disk}}/ 10^{46}})^{1/2}\ pc$ \citep{2008MNRAS.387.1669G}. Thus, the spectral index change depends on the disk luminosity as
\begin{equation}\label{eq10}
\Delta \gamma \simeq \frac{\tau_{\rm He}}{4} \simeq \frac{L_{He}\sigma_\mathrm{T}}{16\pi R_{\rm He}cE_{\rm He}} \simeq0.47\, {\left(\frac{L_{\rm disk}}{\rm 10^{46} erg\,s^{-1}}\right)}^{ \frac{1}{2}}.
\end{equation}

\renewcommand\thefigure{7}
\begin{figure}
\centering
 \includegraphics[height=8cm,width=10cm]{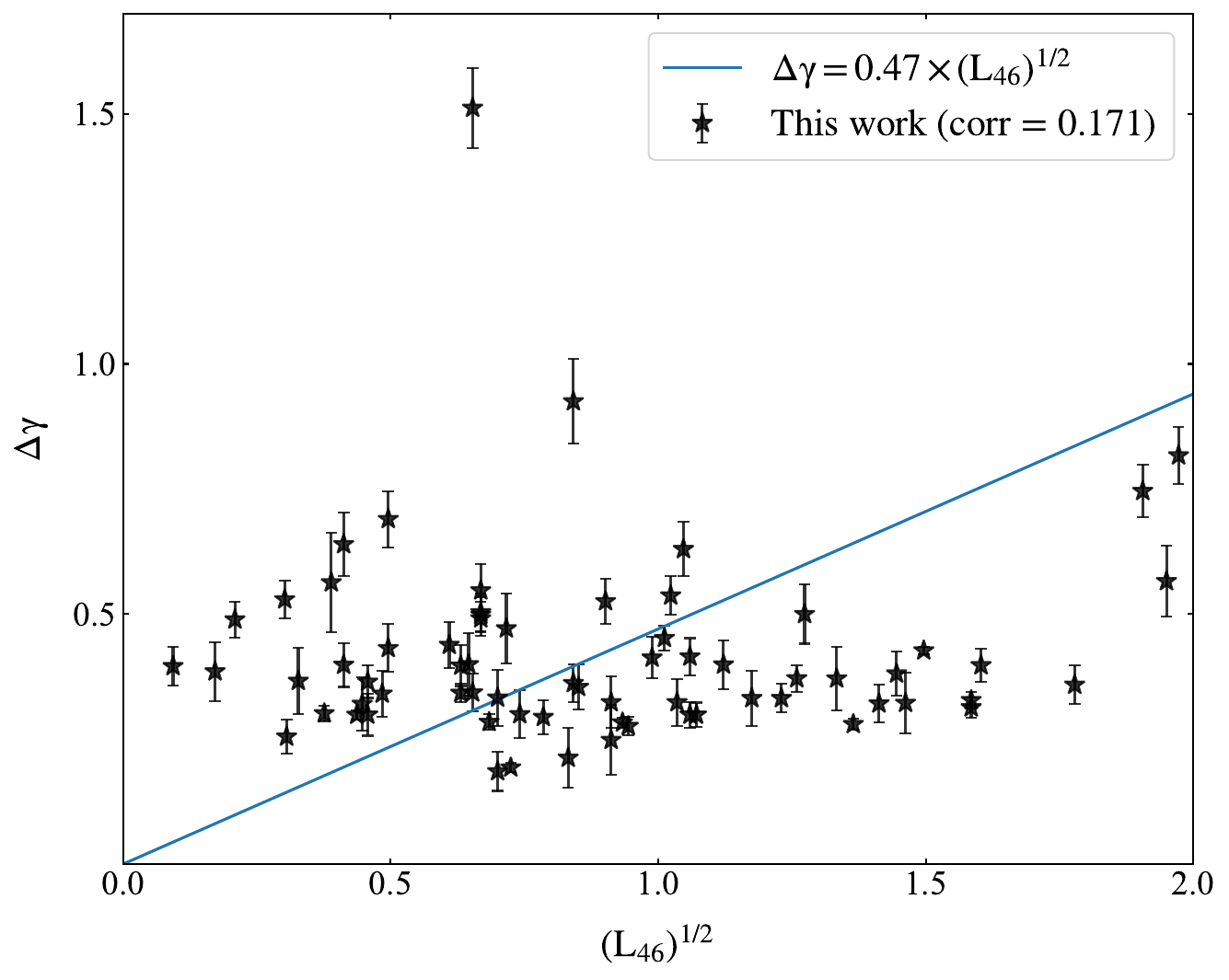}
\caption{The relationship between the observed spectral index change $\rm \Delta \gamma$ and the disk luminosity $\rm {(L_{46})}^{1/2}$. The blue line is the fitted line of Equation \ref{eq10}. The Spearman correlation coefficient between the two parameters is 0.171.   }
 \label{fig:fig7}
\end{figure}

\citet[hereafter ZH20]{2020ApJ...897...10Z} and \citet[hereafter CH24]{2024ApJS..271...20C} systematically collected the BLR luminosity and Doppler factors for a sample of bright FSRQs. Out of the 87 FSRQs we analyzed, 77 are drawn from their original datasets. The parameters collected of 77 samples are summarized in Table \ref{tab:5}. Figure \ref{fig:fig7} illustrates the relationship between the observed spectral index change and the accretion disk luminosity, with the blue line representing the linear fit defined by Equation \ref{eq10}. 
Notably, a majority of data points exhibit a significant deviation from this fitted line, and the correlation between $\rm \Delta \gamma$ and $\rm {(L_{46})}^{1/2}$ is weak, as indicated by a Spearman correlation coefficient of 0.171. This suggests that photon pair production in the BLR is an unlikely mechanism for explaining the origin of the GeV break.

\subsubsection{Radiative losses resulting in an electron spectrum break}

The break of the intrinsic electron spectrum could be a possible cause for the GeV spectral break. Previous observations, such as those by \cite{2010ApJ...710.1271A} and \cite{2010ApJ...721.1383A}, found changes in photon index of $\rm \Delta\gamma > 1$, which exceeded the expected value for a ``cooling break" caused by radiation losses. In the scenario of  slow cooling, the change in electron spectral index $\rm \Delta p$ due to radiation losses is expected to be 1 (e.g. \citealt{2013ApJ...768...54B, 2018MNRAS.478.3855Z}), corresponding to a change in the photon index $\rm \Delta\gamma = 0.5$ . Present BPL fitting shows a change in spectral indexes $\rm \Delta\gamma = 0.45\pm0.19$, which is consistent with the expected change due to radiative losses. This provides possible evidence for the association between the GeV spectral break and the intrinsic electron spectrum break caused by radiation losses.

In the leptonic model, the gamma-ray emission from FSRQs is generally attributed to the inverse Compton scattering of external photon fields by relativistic electrons in the jet. The external photon fields mainly include the BLR and the dusty molecular torus (DT). The electron energy is primarily dissipated through Comptonization with BLR and DT photons  \citep{2009ApJ...704...38S}. 
The BLR is donated by Ly$\alpha$ photon; the typical photon energy is $\sim$10.2 eV (e.g. \citealt{2010ApJ...721.1383A, 2013ApJ...771L...4C}); and the typical size is $\rm R_{BLR} \sim$0.1 pc \citep{2009ApJ...704...38S}. 
The DT is donated by Infrared photons with a characteristic temperature of about 1000K; the typical photon energy is $\sim$0.3 eV; and the typical size is $\rm R_{DT}\sim$4 pc \citep{2009ApJ...704...38S}. 

The GeV break energy and the intrinsic electron spectrum break Lorentz factor satisfy $\rm E_{\mathrm{b}}^{\prime}\simeq\gamma_{b}^{2}\Gamma^{2}E_{\mathrm{ext}}$ \citep{2009ApJ...704...38S}. Where $E_{\rm ext}$ is the typical energy of the external photon field in the rest frame, and $\Gamma$ is the bulk Lorentz factor of the jet.  In extreme relativistic jets of blazars, the Doppler factor $\rm \delta \approx \Gamma$ (e.g. \citealt{2014Natur.515..376G, 2016MNRAS.461.1862K, 2020ApJ...897...10Z, 2021ApJ...915...59Z}). The GeV break at $\sim$ 2.9 GeV suggests that the break in the electron spectrum occurs at a Lorentz factor on the order of $\rm \sim 10^3$ (See Table \ref{tab:5}). The electron break energy corresponds to the critical energy of electron escape and cooling equilibrium in slow cooling, estimated by \citep{2009ApJ...704...38S}
\begin{equation}
\gamma_b\simeq \frac{m_ec^2\Gamma}{\sigma_Tu_{\mathrm{ext}}^{\prime}R}\mathrm.
\end{equation}\label{eq11}
Where $\rm u_{\mathrm{ext}}^{\prime}$ is the energy density of the external photon field in the comoving frame of the jet, and it depends on the location of the emission region as (e.g. \citealt{2009ApJ...704...38S,2012ApJ...754..114H, 2017ApJS..228....1Z})
\begin{equation}
\begin{aligned}
u_{\mathrm{ext}}^{\prime}&=u_{\mathrm{BLR}}^{\prime}+u_{\mathrm{DT}}^{\prime}\\
&=\frac{\xi_{\mathrm{BLR}}L_{\mathrm{disk}}\Gamma^2}{4\pi R_{\mathrm{BLR}}^2c[1+(R/R_{\mathrm{BLR}})^3]}+\frac{\xi_{\mathrm{DT}}L_{\mathrm{disk}}\Gamma^2}{4\pi R_{\mathrm{DT}}^2c[1+(R/R_{\mathrm{DT}})^4]}
\end{aligned}
\end{equation}\label{eq12}
Here, $\rm \xi_{\mathrm{BLR}}$ and $\rm \xi_{\mathrm{DT}}$ is the covering factor of BLR and DT, respectively, with a typical value of 0.1 \citep{2008MNRAS.387.1669G}. The accretion disk luminosity is fixed at $\rm 7.59\times 10^{45}\, erg\,s^{-1}$, while the jet Lorentz factor is set to 26.52, which correspond to the mean value observed in 87 bright FSRQs (See Table \ref{tab:5}). The radius of the DT is modeled as a function of the accretion disk luminosity, following the relation $\rm R_{\mathrm{DT}}\simeq4\left(\frac{L_{\mathrm{disk}}}{10^{46}\mathrm{~erg~s}^{-1}}\right)^{1/2}\left(\frac T{10^3\mathrm{~K}}\right)^{-2.6}\text{ pc }$ \citep{2009ApJ...704...38S}.

We estimate how the energy density of the external photon field, $\rm u_{\mathrm{ext}}^{\prime}$, and the electron spectrum break energy, $\rm \gamma_b$, evolves with the radius of the emission region $\rm R$. For more details, see Figure \ref{fig:fig8}(a). The figure shows that when the emission region radius is less than 1 pc, the external photon field is mainly from BLR photons. When the radius exceeds this, it is dominated by photons from the dusty torus.
In both the BLR-dominated region (Figure \ref{fig:fig8}(b)) and the DT-dominated region (Figure \ref{fig:fig8}(c)), we simulate the energy of the photon spectral break $\rm E_b^{\prime}$  as a function of the emission region radius $\rm R$. The results indicate that if the emission region is less than 1 pc, a spectral break at a few GeV is unlikely. However, if the emission region extends beyond the DT ($\rm \gtrsim 8\, pc$), a break at ~2.9 GeV can occur.

This indicates that the gamma-ray emission region is located outside the DT, far away from the central black hole. This location is highly consistent with the gamma-ray emission region given in \cite{2024ApJS..271...27Z}. 
This hypothesis is supported by the idea that the external photon field of FSRQs is mainly made up of infrared photons from the DT (e.g., \citealt{2014ApJS..215....5K, 2018A&A...616A..63A, 2020ApJS..248...27T}). 
It also helps explain why there is no gamma-ray absorption by the BLR for photons below 10 GeV (see Section 4.3.2).

Our simple simulation shows the potential for radiative cooling to cause the GeV spectral break under certain conditions. However, more observational and theoretical evidence is needed to confirm this scenario.

\renewcommand\thefigure{8}
\begin{figure}
\centering
 \includegraphics[height=7.5cm,width=10cm]{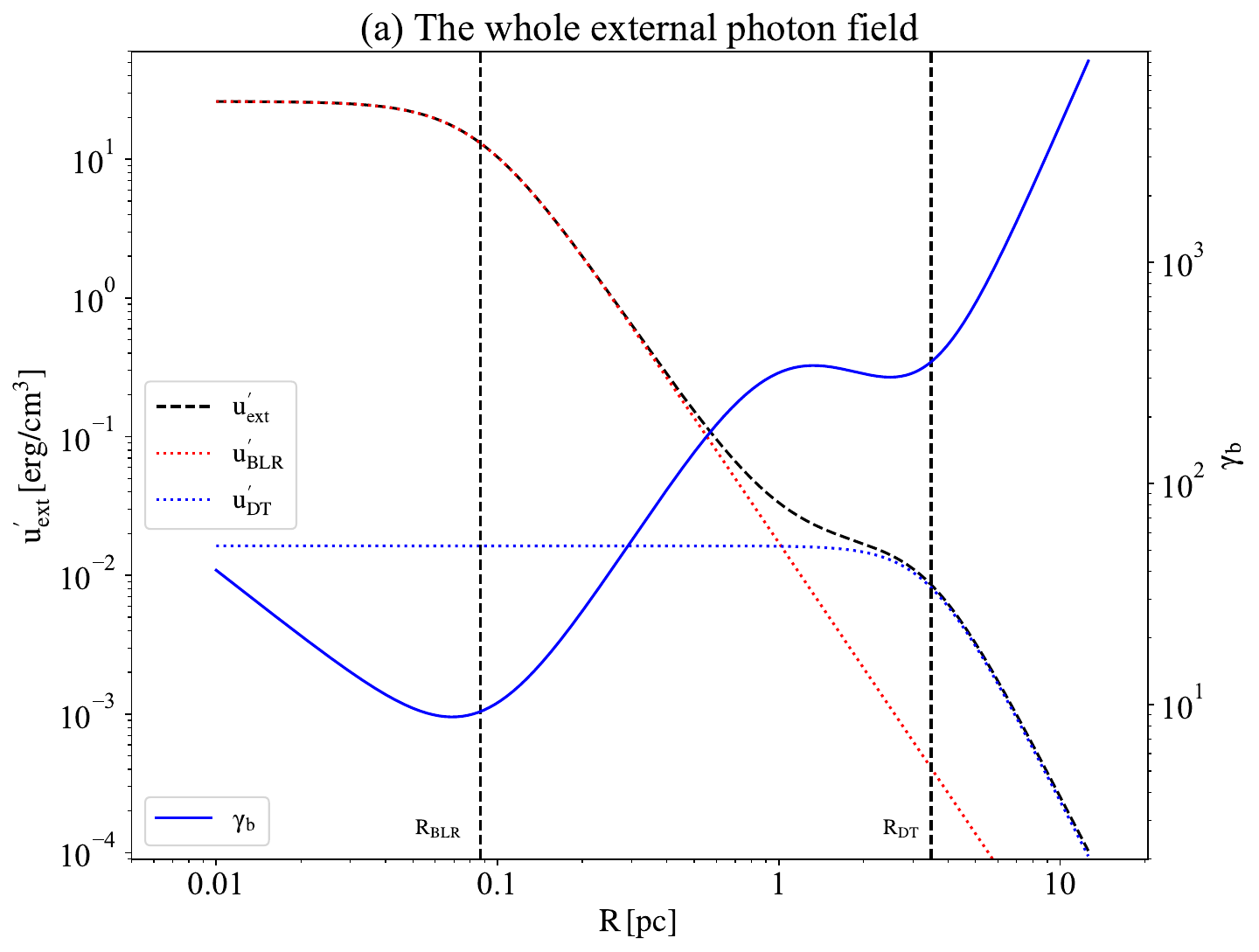}
  \includegraphics[height=7.5cm,width=10cm]{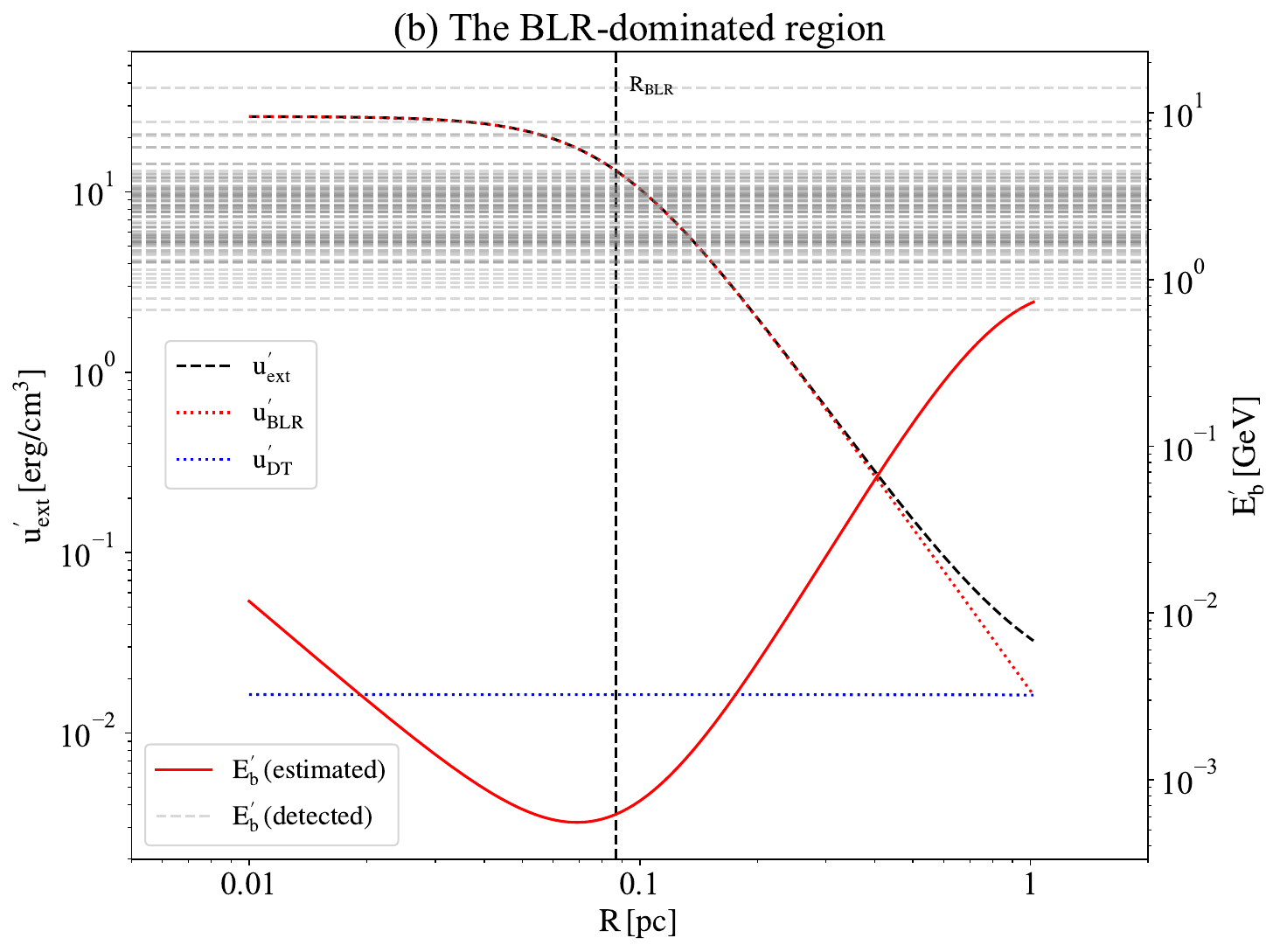}
   \includegraphics[height=7.5cm,width=10cm]{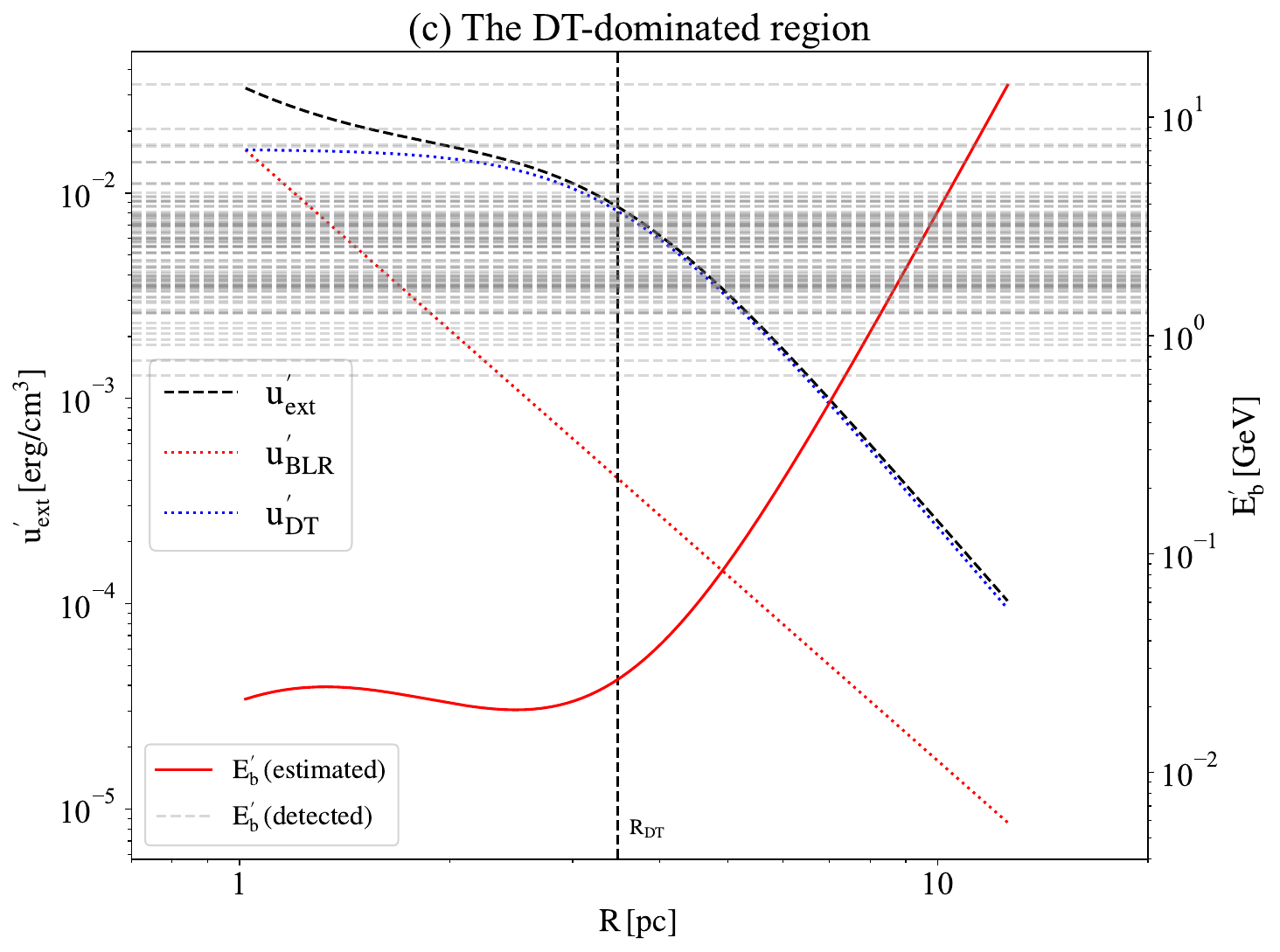}
\caption{The relationship between the emission region radius $\rm R$, the energy density of the external photon field $\rm u_{\mathrm{ext}}^{\prime}$, the electron spectrum break energy $\rm \gamma_b$, and the GeV spectral break energy $\rm E_b^{\prime}$ is shown in: (a) the whole external photon field, (b) the BLR-dominated region, and (c) the DT-dominated region. The grey dashed line represents the GeV break energy detected in context.}
 \label{fig:fig8}
\end{figure}

\subsubsection{KN effect}

Another possibility is that the jet electrons scatter the external photon field in the KN region. 
 In the KN regime, scattering of $\rm Ly\alpha$ photons does indeed lead to spectral softening around $\sim$2 GeV, but the KN effect alone is insufficient to produce a sharp break \citep{2010ApJ...721.1383A}. Cooling effects (e.g. \citealt{2002ApJ...568L..81D, 2005MNRAS.363..954M, 2009ApJ...704...38S}), mixed photon fields (e.g. \citealt{2010ApJ...714L.303F,2013ApJ...771L...4C,2016A&A...589A..96H, 2021MNRAS.502.5875K}), or an intrinsic break in the electron spectrum (e.g. \citealt{2010ApJ...714L.303F, 2021MNRAS.502.5875K}) are introduced to improve the fitting. The GeV break may have a mixed origin. The GeV break we detected is below the maximum photon energy expected from the KN effect, so the KN effect cannot be ruled out. Identifying the multiple origins is beyond the scope of this work and requires further investigation in future studies.

\renewcommand\thetable{5}
\begin{table}
\begin{center}

\caption{Collected parameters and estimated electron break Lorentz factor}
\label{tab:5}
\begin{adjustwidth}{2.2cm}{-5cm}
\resizebox{12cm}{!}{
\begin{tabular}{cccccc} \hline
{Source Name}& Doppler factor & $\rm log_{10}L_{BLR}$ & References& $\rm \gamma_b (BLR)$&$\rm \gamma_b (DT)$ \\
\normalsize(1) & \normalsize(2) & \normalsize(3) &\normalsize(4) & \normalsize(5)  & \normalsize(6) \\
\hline
3C 454.3                    	&	36.02	&	45.35 	&	CH24	&	357 	&	2082 	\\
PKS 1510-089                	&	19.64	&	44.72 	&	CH24	&	656 	&	3828 	\\
3C 279                      	&	61.33	&	44.28 	&	CH24	&	194 	&	1129 	\\
CTA 102                     	&	35.61	&	45.27 	&	CH24	&	341 	&	1988 	\\
PKS 1424-41                 	&	84.88	&	44.94 	&	CH24	&	162 	&	943 	\\
4C +01.02                   	&	37.56	&	45.62 	&	CH24	&	474 	&	2764 	\\
PKS 1502+106                	&	43.09	&	45.40 	&	CH24	&	375 	&	2188 	\\
4C +38.41                   	&	26.46	&	45.69 	&	CH24	&	667 	&	3887 	\\
PKS 0454-234                	&	56.66	&	44.60 	&	CH24	&	187 	&	1089 	\\
4C +21.35                   	&	8.08	&	45.18 	&	CH24	&	2139 	&	12470 	\\
3C 273                      	&	1.53	&	45.41 	&	CH24	&	7311 	&	42631 	\\
Ton 599                     	&	30.16	&	44.67 	&	CH24	&	438 	&	2554 	\\
PKS 0402-362                	&	15.75	&	45.62 	&	CH24	&	1002 	&	5843 	\\
B2 1520+31                  	&	44.44	&	45.01 	&	CH24	&	461 	&	2690 	\\
4C +28.07                   	&	20.4	&	45.40 	&	CH24	&	650 	&	3789 	\\

$\cdots$&$\cdots$&$\cdots$&$\cdots$&$\cdots$&$\cdots$\\

\hline
Average&26.52 &44.88&&1018 &5933\\
\hline
\end{tabular}}
\end{adjustwidth}
\end{center}
{\footnotesize{}}
\end{table}

\section{Conclusion} \label{sec:conclusion}

We present a survey of the GeV spectral breaks in Fermi FSRQs within the energy range of 0.1-10 GeV. A total of 755 FSRQs with known redshifts at high latitudes ($\rm |b| \geq 10^{\circ}$) are selected. Based on Fermi-LAT P8R3 data, both the default spectral models provided by the Fermi team and the broken power-law (BPL) model are used to fit the 15-year-averaged GeV gamma-ray spectra of FSRQs. We provided spectral fitting results of 87 bright FSRQs and evaluated the goodness of fit for both models. The fitting results indicate:

\begin{itemize}	
 \item [1.]
The FSRQ population shows a comparable preference for both the BPL and LP models in gamma-ray spectral modeling, with the fitted lines from both models highly overlapping. Clustering analysis suggests that the BPL-preferred FSRQs and the LP-preferred FSRQs belong to the same sample category. This provides evidence for the equivalence of the BPL and LP models in modeling the 0.1-10 GeV gamma-ray spectra of FSRQs, indicating that the LP model may be a smooth mathematical approximation of the BPL model.
 \item [2.]
GeV spectral breaks are commonly observed in FSRQs. In the rest frame, the break energy is $\rm 2.90\pm1.92$ GeV, which is far from the characteristic absorption energy of the strong emission lines in the BLR. There is no significant correlation between the change in spectral break $\rm \Delta \gamma$ and the luminosity of the accretion disk $\rm L_{disk}$. Therefore, photon pair production in the BLR is unlikely to account for the origin of the GeV spectral break.
 \item [3.] 
 The change in photon index, $\rm \Delta \gamma = 0.45 \pm 0.19$, is consistent with a cooling break in the internal electron spectrum caused by radiative losses. The GeV spectral break suggests that the Lorentz factor for the cooling break is on the order of $\sim 10^3$. 
 When the gamma-ray emission region is located far from the black hole and outside the DT, the Comptonization of external photon fields can effectively cool electrons at this energy, resulting in a photon spectral break at several GeV.
\end{itemize}	

Our analysis focuses on the spectral break in the 0.1-10 GeV energy range. Based on the 15-year-average spectra, we primarily determine the long-term behavior of FSRQ spectra. In fact, some studies suggest that the GeV break observed in 3C 454.3 is quite stable, with the break energy remaining essentially unchanged even during flaring periods \citep{2010ApJ...721.1383A}. However, the preference of gamma-ray spectra for the BPL and LP models may shift over time for some FSRQs \citep{2012ApJ...761....2H,2014MNRAS.441.3591H}. Further investigation is needed to study the evolution of the GeV break phenomenon in the time series. 

The phenomenon of the GeV spectral break is not exclusive to FSRQs but is also observed in some low-synchrotron-peaked BL Lacs \citep{2010ApJ...721.1383A}. Investigating the GeV break in low-synchrotron-peaked BL Lacs will be an important probe into the physical differences in emission regions among different classes of blazars.

This work focuses on the collective behaviour of the FSRQ population. The spectral fitting quality of some sources is poor, with significant uncertainty in parameters such as break energy and photon index. Furthermore, the break of some sources is very slight (with a small $\rm \Delta\gamma$), or the break energy is close to the energy threshold of our analysis. Caution is required when conducting independent studies.

\section*{Acknowledgements}

We thank the anonymous referee for very constructive and helpful comments and suggestions, which greatly helped us to improve our paper. 
We would like to thank the Fermi Science Support Center (FSSC) for the public availability of Fermi-LAT data.
This work is supported by the National Natural Science Foundation of China NSFC 12233006. This work is partially supported by the Yunnan University graduate research innovation fund project (KC-23233826) and the Scientific Research Fund of Education Department of Yunnan Province (2024Y036).
\\

{\it Facility:} Fermi-LAT
\\

{\it Software:}  Fermitools-conda, Fermipy\citep{2017ICRC...35..824W}, Matplotlib\citep{Hunter:2007}, NumPy\citep{ harris2020array}, Astropy \citep{astropy:2022},  Pandas\citep{2022zndo...3509134T}, Seaborn\citep{Waskom2021}, SciPy\citep{2020SciPy-NMeth}, Scikit-learn\citep{scikit-learn}.

\bibliographystyle{aasjournal}
\bibliography{ms}

\renewcommand\thefigure{1}
\begin{figure}[H]
\centering
\includegraphics[height=23cm,width=18cm]{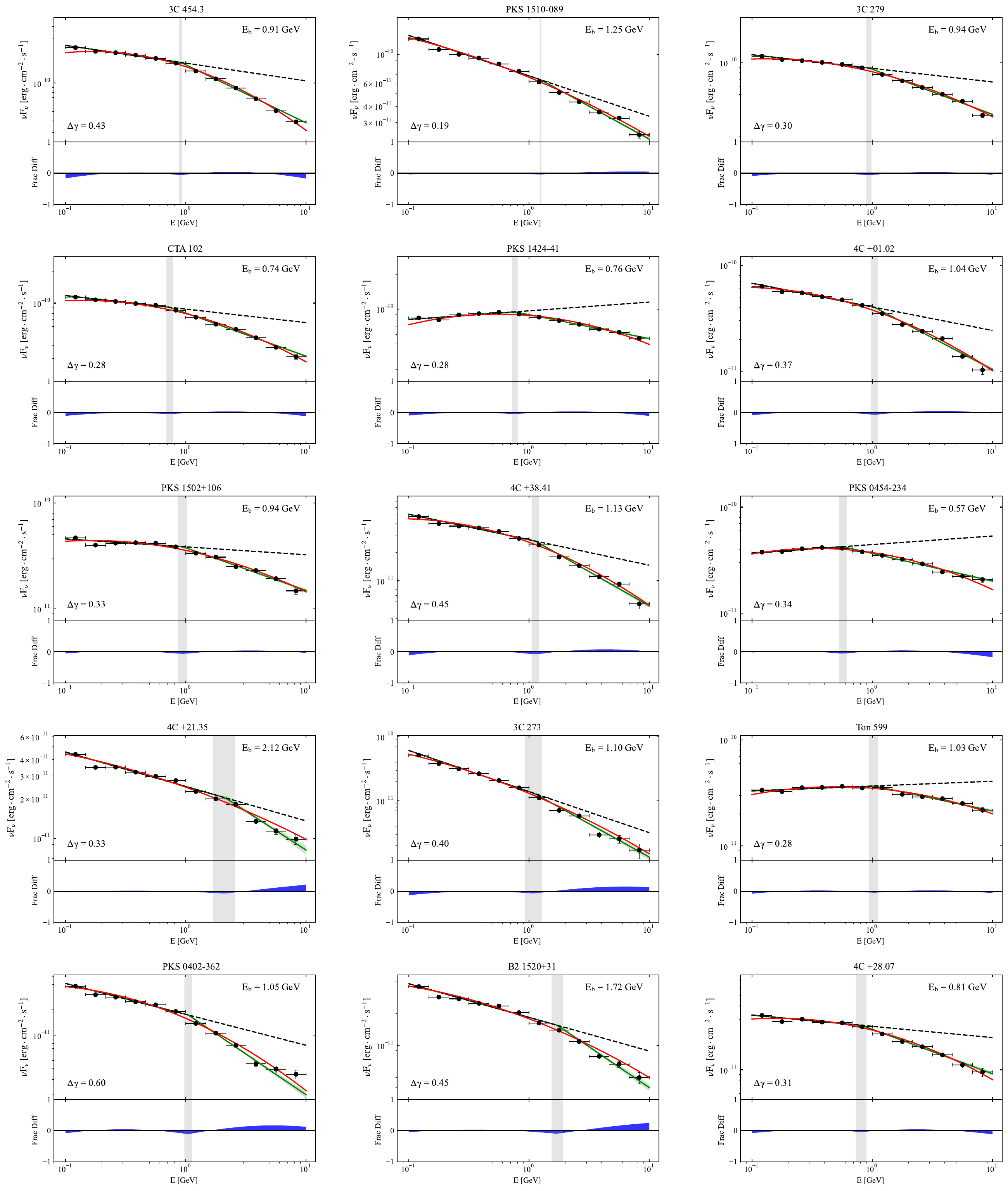}
 \caption{The fitting results of the 15-year average gamma-ray spectra for 87 bright FSRQs. Red line and red area: Best fit line and its 1-$\sigma$ confidence region for LP model; Green line and green area: Best fit line and its 1-$\sigma$ confidence region for the BPL model; Black dashed line: Extension line for the low-energy power-law component of the BPL model; Black (blue) dots: Real SED data significant at 5 $\sigma$ (3 $\sigma$); Gray area: Spectrum break energy range; Blue area: Fractional difference between the two model fit lines.}
 \label{fig:fig1}
\end{figure}

\renewcommand\thefigure{1}
\begin{figure}[H]
\centering
\includegraphics[height=23cm,width=18cm]{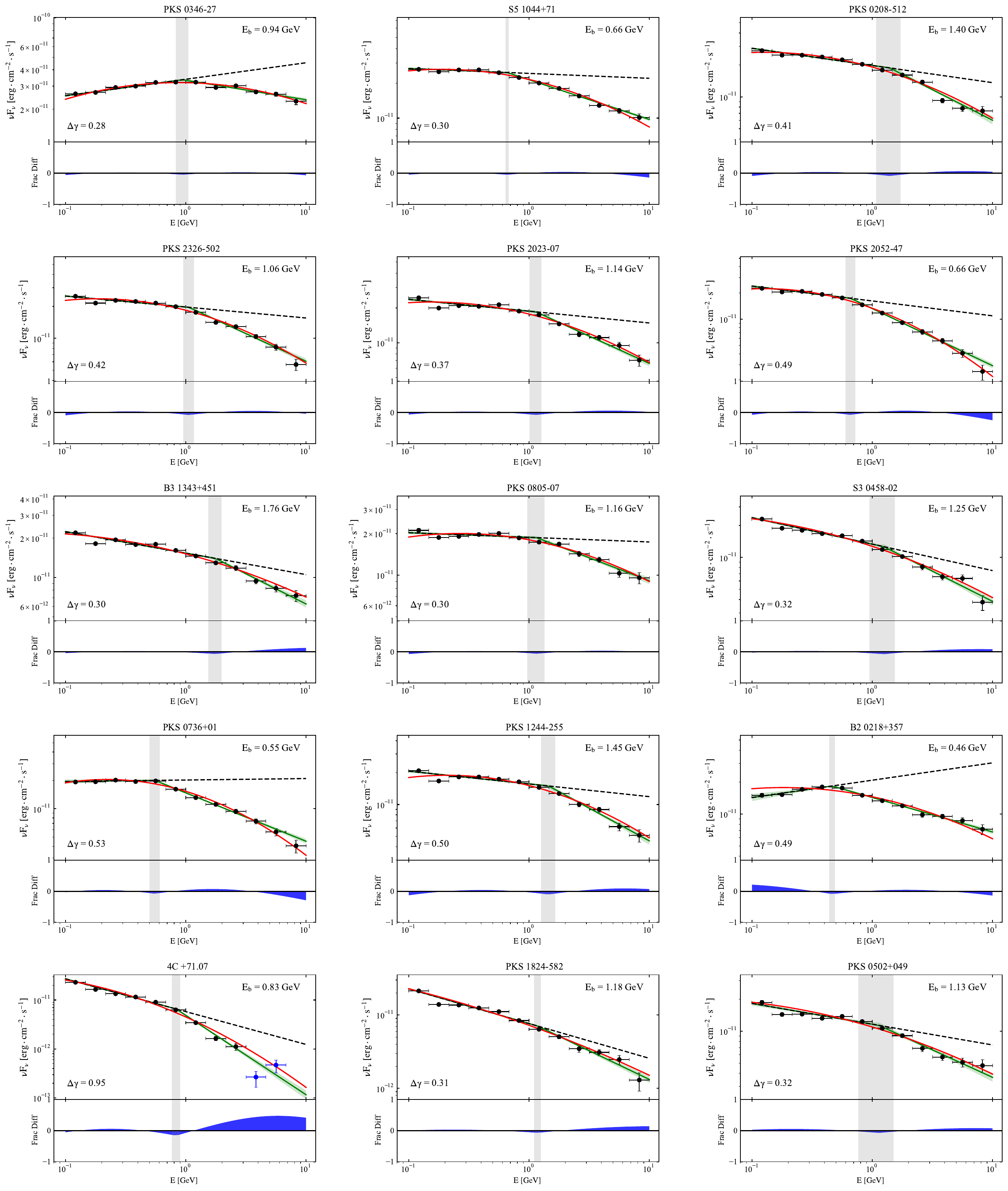}
 \caption{Continued}
\end{figure}

\renewcommand\thefigure{1}
\begin{figure}[H]
\centering
\includegraphics[height=23cm,width=18cm]{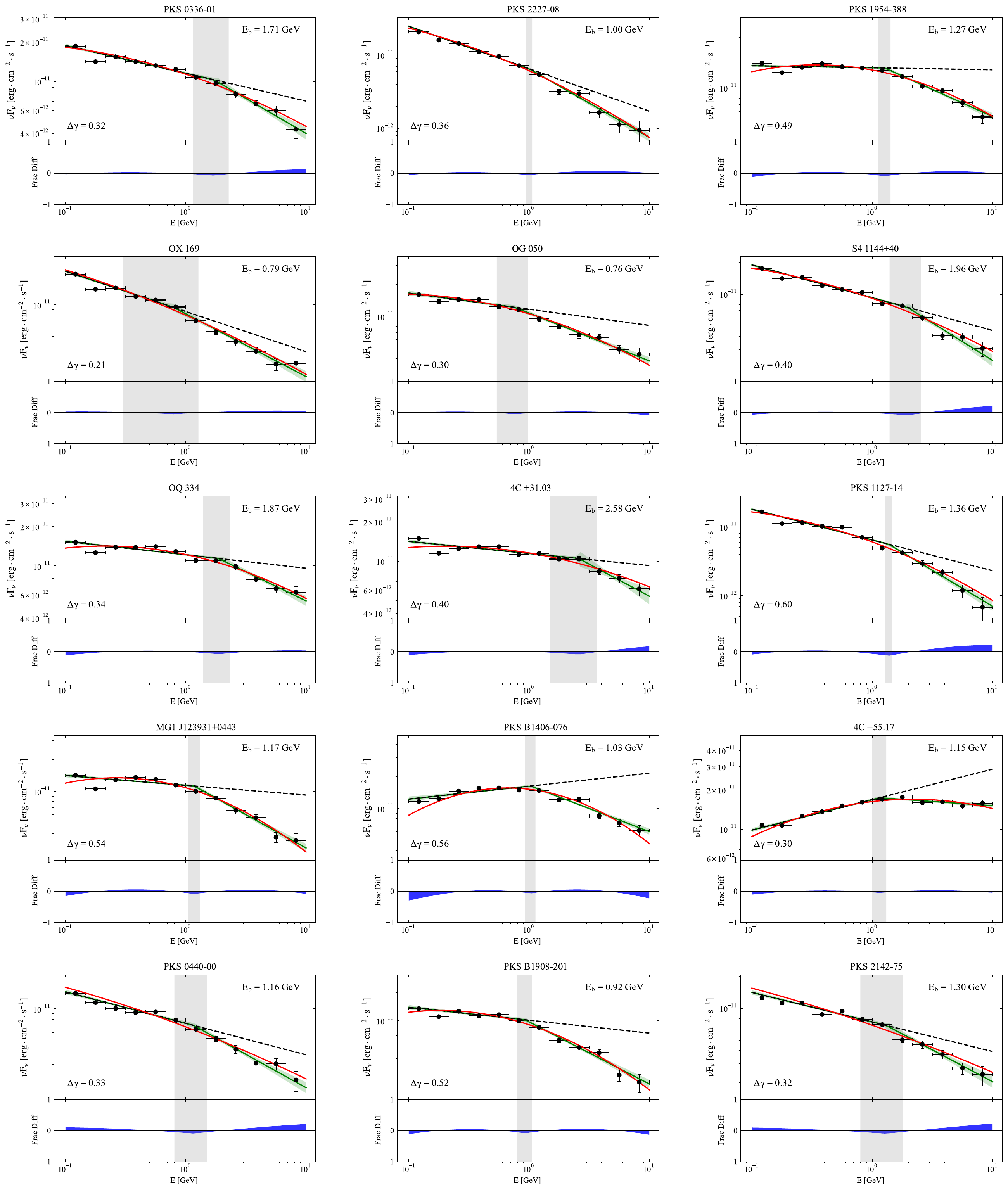}
 \caption{Continued}
\end{figure}

\renewcommand\thefigure{1}
\begin{figure}[H]
\centering
\includegraphics[height=23cm,width=18cm]{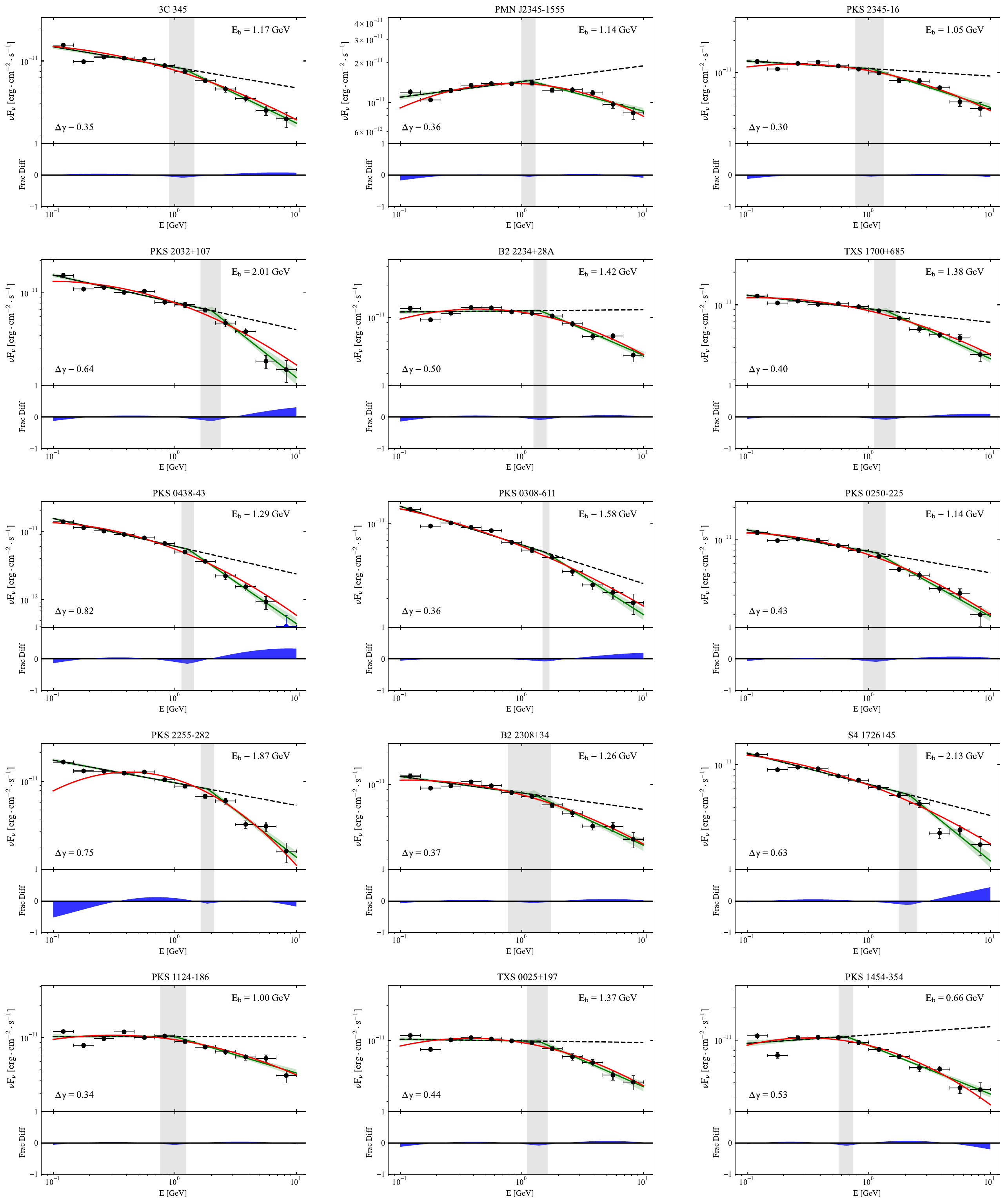}
 \caption{Continued}
\end{figure}

\renewcommand\thefigure{1}
\begin{figure}[H]
\centering
\includegraphics[height=23cm,width=18cm]{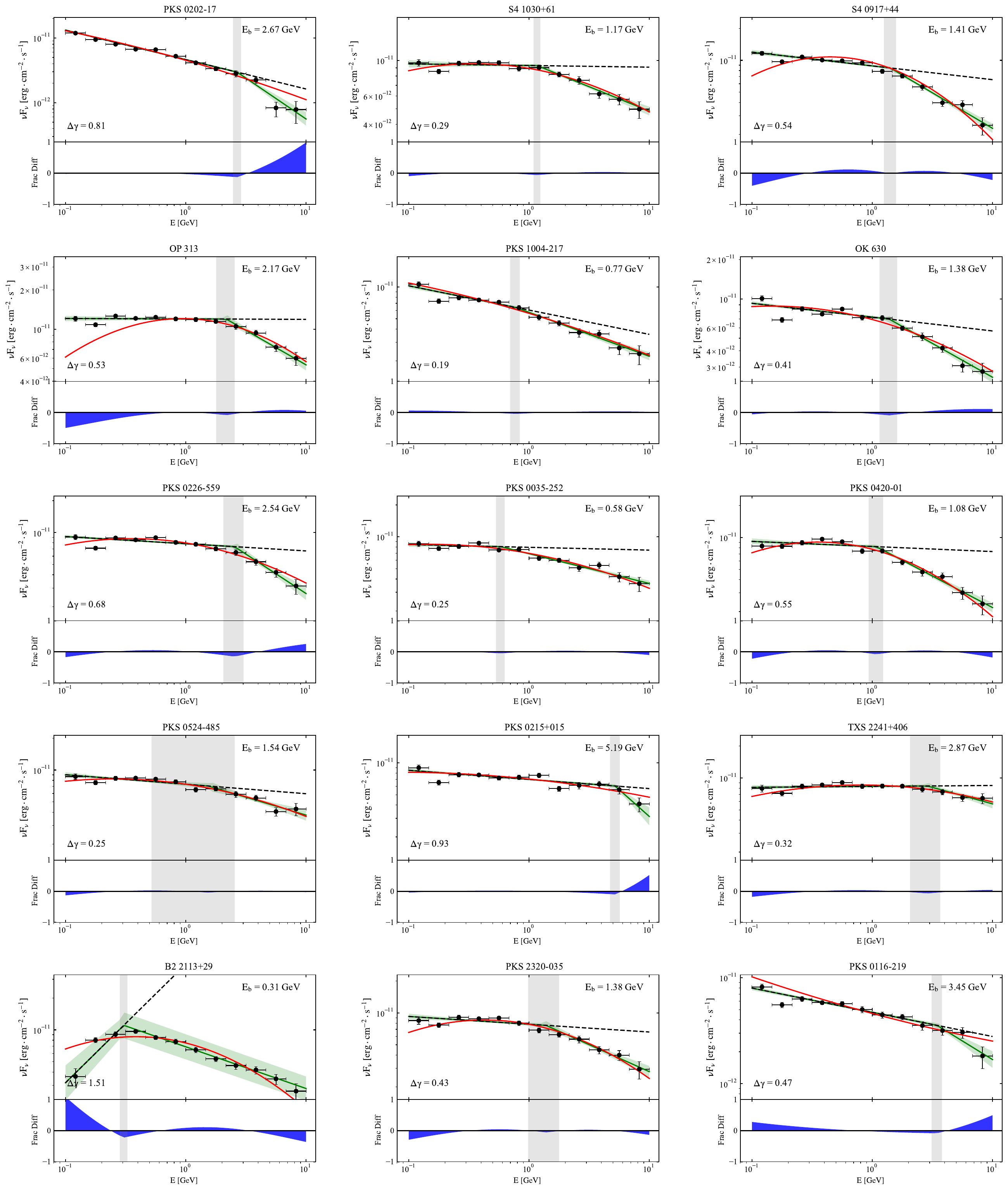}
 \caption{Continued}
\end{figure}

\renewcommand\thefigure{1}
\begin{figure}[H]
\centering
\includegraphics[height=18.5cm,width=18cm]{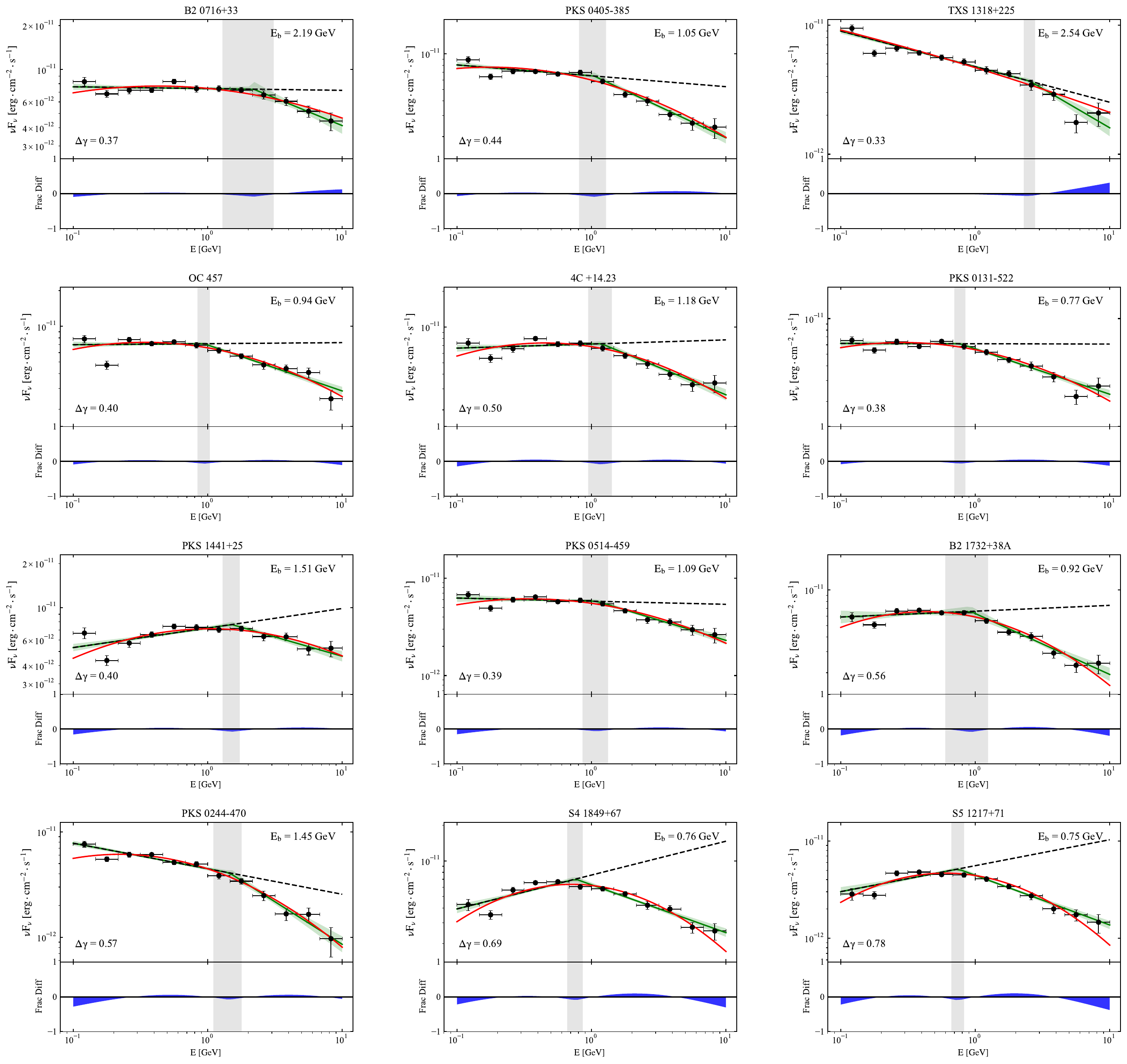}
 \caption{Continued}
\end{figure}

\newpage

\appendix              

\section{Comparison of LP-preferred FSRQs and the BPL-preferred FSRQs} \label{sect:appendix a}

\renewcommand\thefigure{A1}
\begin{figure*}
\centering
\includegraphics[height=18cm,width=18cm]{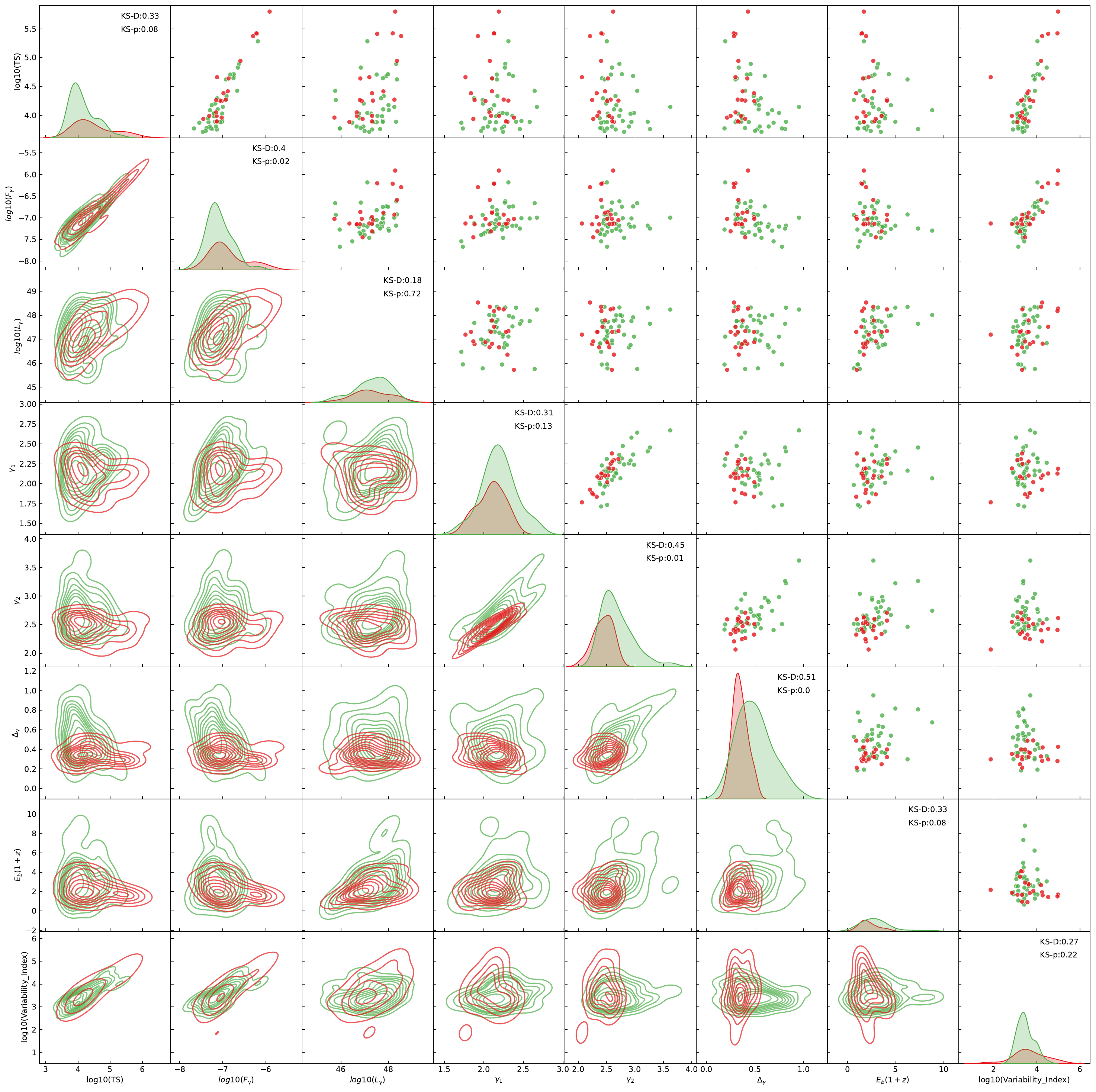}

 \caption{Parameter distribution of the LP-preferred FSRQs (red points) and BPL-preferred FSRQs (green points).The text in the figure indicates the KS test statistics (D) and p-values (p) in the parameter space.}
 \label{fig:figA1}
\end{figure*}

We evaluate the fitting quality of the 0.1-10 GeV gamma-ray spectra for 87 bright FSRQs using the Akaike Information Criterion (AIC). By comparing the differences in AIC between the BPL model and the default Fermi spectral model (LP), we found that FSRQs exhibit different preferences for these two models. Twenty FSRQs show significantly better fitting results when fitted with the BPL model (referred to as BPL-preferred FSRQs), while 39 sources exhibit better fits when fitted with the LP model (referred to as LP-preferred FSRQs). 
These two types of sources show a strong preference for the BPL or LP models in the AIC test, but the fitted lines from both models largely overlap.

To explore potential intrinsic physical differences between these two sample types, we collect physical parameters for parameter distribution analysis. Specifically, we gather three parameters from Fermi default spectral fitting: TS value, gamma-ray luminosity $\rm l_\gamma$, and gamma-ray flux $\rm F_\gamma$. From the BPL model fitting, we obtain four parameters: low-energy photon index $\gamma_1$ and high-energy photon index $\gamma_2$, change in photon index $\rm \Delta \gamma$ and spectral break energies in the rest frame $\rm E_b^{\prime}$. Additionally, we collect the gamma-ray band variability index $\rm VI$ from 4FGL-DR4 \citep{2023arXiv230712546B}. 
Figure \ref{fig:figA1} illustrates the distribution of BPL-like FSRQs (red points) and LP-like FSRQs (green points) in an eight-dimensional parameter space. We find that BPL-like FSRQs exhibit smaller photon spectral indexes and smaller changes in photon spectral indexes. For other parameters, the distributions of the two sample types highly overlap, and no significant differences were observed.

We used the Kolmogorov-Smirnov (KS) test to explore the distribution of various parameters for the two groups of FSRQs. The corresponding KS statistics (D) and p-values (p) are shown on the diagonal of Figure \ref{fig:figA1}. At a significance level of 0.01, the hypothesis that the two sample groups follow the same distribution cannot be rejected for any parameter except for the change in photon index $\rm \Delta \gamma$ .

The Gaussian Mixture Model (GMM) is a probabilistic model that assumes the dataset is composed of multiple Gaussian distributions. It works by estimating the parameters of each Gaussian distribution to model and analyze the data, commonly used for tasks such as clustering analysis, density estimation, and anomaly detection. 
Based on the \emph{scikit-learn}, we construct a GMM model and performed clustering analysis on an eight-dimensional parameter dataset containing BPL-preferred FSRQs and LP-preferred FSRQs \citep{scikit-learn}. The fitting of GMM with different numbers of clusters ($\rm n\_clusters$) is conducted. the goodness the GMM model is evaluated using the Bayesian Information Criterion (BIC).
The evolution of the model's BIC values with the number of clusters is shown in Figure \ref{fig:figA2}. From the figure, it can be seen that the BIC value is minimized when considering only one cluster, indicating the optimal model. This provides evidence that BPL-preferred FSRQs and LP-preferred FSRQs belong to the same type and have consistent origins of gamma-ray emissions.

\renewcommand\thefigure{A2}
\begin{figure*}
\centering
\includegraphics[height=6cm,width=8cm]{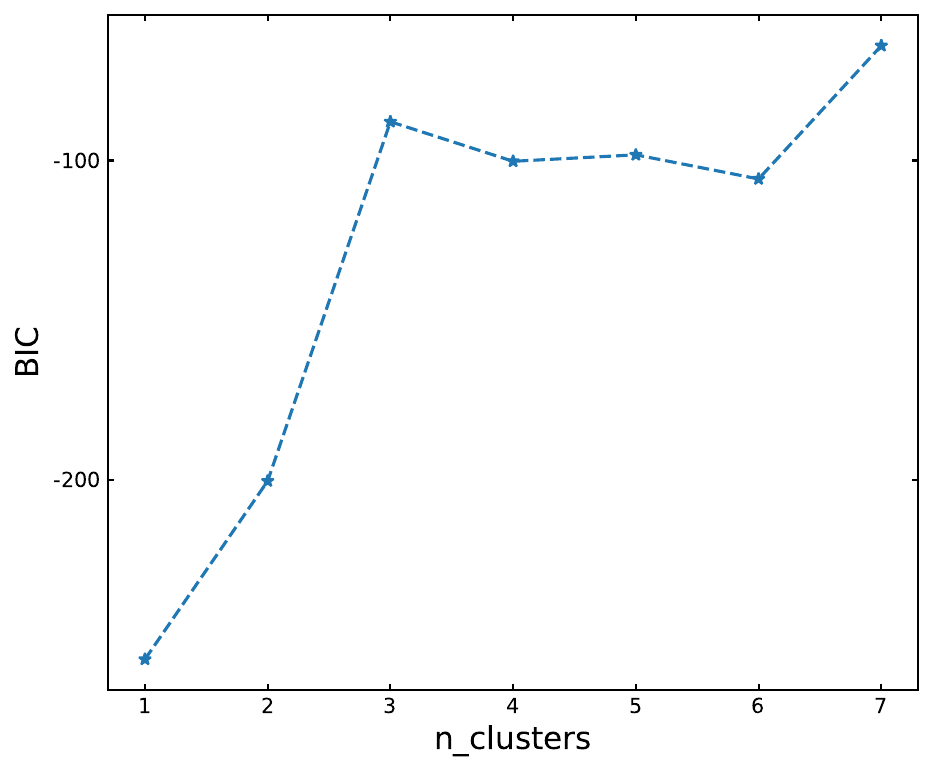}

 \caption{BIC after GMM clustering with different number of clusters ($\rm n\_clusters$)}
 \label{fig:figA2}
\end{figure*}

\end{document}